\documentclass[runningheads]{svmult}
\usepackage{makeidx}   
\usepackage{graphicx}  
\usepackage{subeqnar}  
\usepackage{multicol}  
\usepackage{physprbb}  
\makeindex             

\newcommand{\greeksym}[1]{{\usefont{U}{psy}{m}{n}#1}}
\newcommand{\umu}{\mbox{\greeksym{m}}}

\begin{document}

{

\def\dofig{1}
\nonstopmode

\newcommand{\FIG}[3]{%
  \begin{figure}[th!]
  \begin{center}
  \ifnum\dofig=1
  \includegraphics[height=#2\textwidth]{mhd01_nordlund_#1.eps}
  \else
  \vspace*{#2\textwidth}
  \fi
  \end{center}
  \caption[]{#3}
  \label{#1.fig}
  \end{figure}}
\newcommand{\FIGG}[4]{%
  \begin{figure}[th!]
  \ifnum\dofig=1
  \centerline{%
  \includegraphics[height=#3\textwidth]{mhd01_nordlund_#1.eps}%
  \includegraphics[height=#3\textwidth]{mhd01_nordlund_#2.eps}}
  \else
  \vspace*{#3\textwidth}
  \fi
  \caption[]{#4}
  \label{#1.fig}
  \end{figure}}
\newcommand{\FIGGG}[5]{%
  \begin{figure}[th!]
  \begin{center}
  \ifnum\dofig=1
  \includegraphics[height=#4\textwidth]{mhd01_nordlund_#1.eps}
  \includegraphics[height=#4\textwidth]{mhd01_nordlund_#2.eps}
  \includegraphics[height=#4\textwidth]{mhd01_nordlund_#3.eps}
  \else
  \vspace*{#4\textwidth}
  \vspace*{#4\textwidth}
  \vspace*{#4\textwidth}
  \fi
  \end{center}
  \caption[]{#5}
  \label{#1.fig}
  \end{figure}}
\newcommand{\FIGGGG}[6]{%
  \begin{figure}[th!]
  \begin{center}
  \ifnum\dofig=1
  \includegraphics[height=#5\textwidth]{mhd01_nordlund_#1.eps}
  \includegraphics[height=#5\textwidth]{mhd01_nordlund_#2.eps}\\
  \includegraphics[height=#5\textwidth]{mhd01_nordlund_#3.eps}
  \includegraphics[height=#5\textwidth]{mhd01_nordlund_#4.eps}
  \else
  \vspace*{#5\textwidth}
  \vspace*{#5\textwidth}
  \fi
  \end{center}
  \caption[]{#6}
  \label{#1.fig}
  \end{figure}}

\newcommand{\Fig}[1]{Fig.\ \ref{#1.fig}}
\newcommand{\Figs}[2]{Figs.\ \ref{#1.fig}--\ref{#2.fig}}
\newcommand{\Figure}[1]{Figure \ref{#1.fig}}
\newcommand{\Sect}[1]{Sect.\ \ref{#1.sec}}

\newcommand{\EQ}[1]{\begin{equation}\label{#1.eq}}
\newcommand{\EN}{\end{equation}}
\newcommand{\Eq}[1]{(\ref{#1.eq})}
\newcommand{\Equation}[1]{Equation (\ref{#1.eq})}

\newcommand{\half}{\frac{1}{2}}
\newcommand{\Mach}{{\cal M}}


\title*{Star Formation and the Initial Mass Function\protect{\footnote{%
Invited tutorial,
{\em Simulations of Magnetohydrodynamic Turbulence in Astrophysics: Recent Achievements and Perspectives},
Eds.: E. Falgarone and T. Passot,
Lecture Notes in Physics}}}
\toctitle{Star Formation and the Initial Mass Function}
\titlerunning{Star Formation and the Initial Mass Function}

\author{{\AA}ke Nordlund\inst{1}
        \and Paolo Padoan\inst{2}}
\authorrunning{{\AA}.\ Nordlund \& P.\ Padoan}

\institute{Astronomical Observatory / NBIfAFG and
     Theoretical Astrophysics Center\\
     Juliane Maries Vej 30, DK-2100 Copenhagen \O, Denmark
\and Jet Propulsion Laboratory, 4800 Oak Grove Drive, MS 169-506\\
     California Institute of Technology, Pasadena, CA 91109-8099, USA}

\maketitle              

\begin{abstract}
Supersonic turbulence fragments the
interstellar medium into dense sheets, filaments, cores
and large low density voids, thanks to a complex network
of highly radiative shocks. The turbulence is driven on
large scales, predominantly by supernovae. While on scales of
the order of the
disk thickness the magnetic energy is in approximate
equipartition with the kinetic energy of the turbulence, on
scales of a few pc the turbulent kinetic energy significantly
exceeds the magnetic energy.

The scaling properties of supersonic turbulence are well
described by a new analytical theory, which allows to
predict the structure functions of the density and velocity
distributions in star-forming clouds up to very high order.

The distribution of core masses depends
primarily on the power spectrum of the turbulent flow, and
on the jump conditions for isothermal shocks in a
magnetized gas. For the predicted velocity power spectrum
index $\beta=1.74$, consistent with results of numerical
experiments of supersonic turbulence as well as with Larson's
velocity-size relation, one obtains by scaling arguments
a power law mass distribution of dense cores with a slope
equal to $3/(4-\beta)$ = 1.33, consistent with the slope of
the Salpeter stellar initial mass function (IMF). Results from numerical
simulations confirm this scaling. Both the analytical model for
the stellar IMF and its numerical estimate show that turbulent
fragmentation can also explain the origin of brown dwarfs.
The analytical predictions for the relative abundance of brown dwarfs
are confirmed by the observations.

The main conclusion is that the stellar IMF directly reflects
the mass distribution of prestellar cores, due predominantly
to the process of turbulent fragmentation.


\end{abstract}

\section{Introduction}

Turbulence in the interstellar medium (ISM) of the Milky Way
-- and more generally turbulence in the discs of other
galaxies -- is of crucial importance for both the structure
and evolution of the galaxy.  The importance of turbulence is
both direct, through its influence on the pressure
equilibrium and stratification, and indirect, through its
influence on the star formation process.

The vertical pressure equilibrium and stratification of the
ISM is determined by the level of turbulence, together with
the temperature distribution of the medium (which is in turn
probably tightly coupled to the turbulence), and it is likely
that even the distributions of magnetic fields and cosmic ray
particles, which also contribute to the pressure and
stratification, are integral parts of the same process; it
is unlikely that the near equipartition of the energy content
of turbulence, magnetic fields, and cosmic ray particles is
a mere coincidence.

It has long been realized that turbulence in the interstellar
medium, in particular in the cold, molecular cloud components,
is highly supersonic \cite{%
1979MNRAS.186..479L,%
1981MNRAS.194..809L,%
1982MNRAS.200..159L,%
1990ApJ...359..344F,%
1992A&A...257..715F}.
More recently, it has been realized that the supersonic
nature of the turbulence is a boon, rather than a nuisance,
when trying to understand the properties of the ISM, the
cold molecular clouds, and star formation \cite{%
1995MNRAS.277..377P,%
1999ApJ...526..279P,%
2000ApJ...530..277E}.
It turns out that supersonic turbulence is in many respect
similar to ordinary, subsonic turbulence, and that it thus
has a number of generic, statistical properties.  Much like
ordinary turbulence, its decay time is of the order of the
dynamical time, even in the MHD-case \cite{%
MacLow_Puebla98,%
1998ApSS.261..195M,%
1999ApJ...524..169M,%
Stone+98,%
1999ApJ...526..279P}.
And much like ordinary turbulence, it is characterized by
power law velocity power spectra and structure functions
over an inertial range of scales \cite{%
Boldyrev,%
Boldyrev+1,%
Boldyrev+2}.

An important difference between supersonic and subsonic
turbulence is the distribution of density.  A supersonic
medium is, by definition, highly compressible; on average
its gas pressure $P_g$ is small relative to the dynamic
pressure $\rho u^2$.  As a consequence, a supersonic medium
is characterized by a wide distribution of densities.  A
turbulent and isothermal supersonic medium has
a log-normal density probability distribution (PDF)
\cite{%
1994ApJ...423..681V,%
1997MNRAS.288..145P,%
Passot+98},
with a dispersion of linear density proportional to the
Mach number
\cite{%
1997MNRAS.288..145P,%
1999intu.conf..218N,%
2001ApJ...546..980O}.
Cold molecular clouds
are indeed approximately isothermal, and are known to have
a very intermittent density distribution, consistent with
the properties of isothermal supersonic turbulence \cite{%
1997ApJ...474..730P}.  Deviations from isothermal conditions
are in general of the type where compression leads to even
lower temperatures (effective gas gamma less than unity)
\cite{1998ApJ...504..835S},
resulting in a density PDF skewed towards greater probability
at high densities.  The PDF may be described as a skewed
log-normal, with a high density asymptote that formally tends
to a power law in the limit $T \rightarrow 0$ \cite{%
1998ApJ...504..835S,1999intu.conf..218N}.

Effectively then, supersonic turbulence acts to fragment the
ISM, causing local density enhancements also over a range of
geometrical scales.  Molecular clouds themselves represent relatively
large scale density enhancements, probably caused by the
random convergence of large scale ISM velocity features
\cite{1999ApJ...515..286B,1999ApJ...527..285B,2001ApJ...562..852H}.
Inside molecular clouds smaller scale
turbulence leads to high contrast local density enhancements
in corrugated shocks, intersections of shocks, and in knots
at the intersection of filaments.  Such small scale density
enhancements are `up to grabs' by gravity; if their density
is sufficiently high, relative to their temperature and the
local magnetic field strength, they form pre-stellar cores,
and eventually collapse to form stars.  The decisive
importance of turbulence in this process makes it possible to
predict the distribution of masses of the pre-stellar cores,
and hence the distribution of new borne stars, the initial
mass function (IMF)
\cite{Padoan+Nordlund-IMF,Padoan_etal-IMF}.

The process of star formation is indeed crucial to understand.
Only by understanding star formation, qualitatively and
quantitatively, can we understand galaxy formation.
We need to understand evolution effects to answer questions
such as ``Was star formation different in the Early Universe?''.
We need to understand environmental effects to answer
questions such as ``Do other galaxies have different `Larson laws'?''

We also need to understand star formation to answer questions
related to Gamma-Ray Bursts; e.g., ``Are Very Massive Stars
progenitors of Gamma-Ray Bursts?'', and ``What environment
does the blast wave associated with Gamma-Ray Bursts encounter''?

Finally, we would like to understand star formation as such,
because it is a neat problem -- one that involves supersonic,
selfgravitating MHD turbulence and thus was thought to be
enormously difficult.  With access to supercomputer modeling
the problem has become tractable, and it has
turned out \emph{a posteriori} that it is even partly amenable
to analytical theory.

In the subsequent sections of this tutorial star formation
and turbulence in the interstellar medium is discussed in
more detail.  Section \ref{sne.sec} discusses supernova
driving of the ISM, Section \ref{sst.sec} discusses
properties of supersonic turbulence, Section \ref{stas.sec}
summarizes a new theory of supersonic turbulence, while Section
\ref{imf.sec} discusses star formation and the initial mass
function.  Conclusions are summarized in Section
\ref{concl.sec}.

\section{Supernova driving of the Interstellar Medium}
\label{sne.sec}

With turbulence being of such fundamental importance in
determining the structure and star formation efficiency of the
interstellar medium it is important to understand what its
primary sources are, and what its overall energy budget is.

First, an estimate and lower limit of the energy input needed
to sustain interstellar turbulence is given by Kolmogorov's
scaling expression for the energy transfer rate in a turbulent
cascade \cite{K41},
\EQ{ISMdiss}
  \epsilon \sim \rho U^3/L ~,
\EN
where $U$ and $L$ are velocity amplitudes and length
scales, respectively.  In Kolmogorov's classical theory this
quantity is assumed to be invariant across the inertial range,
and for our purposes this is adequate; subsequent enhancements
of Kolmogorov's theory \cite{SL94} and modifications for
supersonic conditions \cite{Boldyrev+2} would not change the
following estimates significantly.

Observationally, the velocity dispersion in the ISM adheres to
Larson's scaling law,
\EQ{LarsonV}
  U \sim 1\,\mathrm{km\,s}^{-1}\,
  \left({L\over\mathrm{pc}}\right)^{\alpha} ~,
\EN
with $\alpha \approx 0.4$ \cite{1979MNRAS.186..479L,%
1981MNRAS.194..809L,1992A&A...257..715F}, which means that an
estimate based on \Eq{ISMdiss} only depends very weakly on the
scale $L$ on which the estimate is based. On scales $L\sim 1$
kpc, the turbulent velocity dispersion is of the order $U
\sim 10~$km\,s$^{-1}$ \cite{1979MNRAS.186..479L}, which leads
to the estimate $\epsilon \sim
5\,10^{50}$~erg\,kpc$^{-3}$\,Myr, using an average ISM density
$\sim 1.5\,10^{-24}$ g\,cm$^{-3}$ \cite{1990ApJ...365..544B}.

For comparison, the rate of energy input to the ISM from
supernovae is of the order of $10^{53}$~erg\,kpc$^{-3}$\,Myr,
based on a rate of one SN per 70 years in a galactic volume
spanned by a radius of 15 kpc and a disk thickness of 200 pc
\cite{2000MNRAS.315..479D}.  Thus, less than one percent of
the average supernova energy input is necessary to sustain the
turbulent cascade of energy in the ISM.

Two questions come to mind:  1) Is there at all a turbulent
cascade, and 2) is the energy from supernova at all available
for feeding such a cascade?

The answer to the first question is definitely affirmative;
whatever its source, the observed velocity field at large
scales can do nothing but drive a cascade towards smaller
scale, since there is no dissipation mechanism that operates
at such large scales.  As has recently been shown \cite{%
Boldyrev+1,Boldyrev+2}, it makes little difference whether
turbulence is subsonic or supersonic; similar cascades
arise in both cases, only details such as power law
exponents differ.  The predicted scaling of the velocity
dispersion with size is consistent with the observed (Larson's
law) scaling \cite{1979MNRAS.186..479L,1981MNRAS.194..809L,%
1992A&A...257..715F}.  Most of the observed scatter around
the expected scaling (e.g., Fig.\ 9 in \cite{1992A&A...257..715F})
is probably due to cloud-to-cloud variations -- observations
for a single cloud (Polaris) define a remarkably well defined
scaling over more than three orders of magnitude in size
\cite{Ossenkopf+02}.  A complementary piece of evidence for
power law behavior comes from the observed relation between
age difference and spatial separation \cite{2000ApJ...530..277E}.

The answer to the second question is less obvious, but in the
end also affirmative.  One might think that supernova energy
input occurs at small scales, and hence cannot be a source at
large scales for the turbulent cascade.  However, as has been
demonstrated by detailed numerical simulations \cite{%
1999ApJ...514L..99K,Gudiksen99,2000MNRAS.315..479D}, supernovae are
indeed capable of sustaining a turbulent cascade with velocity
dispersions consistent with observed values.  The transfer of
energy to large scales occurs through the expansion of
supernova bubbles and super-bubbles; i.e., via the hot
component of the ISM.  The hot component coexists with cooler
components (or rather a continuous distribution of
temperatures), and the expansion of the hot component into
channels and chimneys creates kinetic energy on large scales,
available for cascading to smaller scales, also in the cooler
components.

The numerical models demonstrate that supernova-driving of the
interstellar medium is a viable and probably dominating
mechanism, at least in the disc of our galaxy. The detailed
numerical models are also broadly consistent with analytical
and semi-analytical models of supernova feedback and turbulent
self-regulation in galactic discs \cite{%
2000MNRAS.317..697E,%
1999ApJ...527..673S}.

One may ask whether other sources of energy input could be
significant.  Winds from hot stars is one candidate that may
contribute \cite{1995ApJ...441..702V,1995ApJ...455..536P}; regions that create
supernovae of type II are likely to also contain hot, early
type stars. Jets from new-borne stars have been mentioned as
an energy input candidate, but it is unlikely to be
significant on scales above a few pc.

Larson's original paper \cite{1979MNRAS.186..479L} lists
irregularities and asymmetries in the rotation curve
as larger scales that fit into a general power law behavior, at
$L \sim 2-10$ kpc, although with a break at $L \sim 1$ kpc.
This could be taken as an indication that
such irregularities, stemming for example from large scale
density waves, could also be a source of driving for the turbulent
cascade.  But the relation could also go the other way;
irregularities on scales of several kpc could be the imprint
of old super-bubbles, stretched in the direction of rotation by
the differential rotation.

In other contexts, such as star-burst galaxies, the balance between
contributions to the driving may be different; kinetic energy input
from collisions or close interactions between galaxies may be an
important energy input channel there, for example.

\section{Turbulent cascade of the Interstellar Medium}
\label{sst.sec}

When regarded as an isolated phenomenon, molecular clouds
have traditionally given rise to concerns about the source
of their turbulence, their life times, and about their
support against gravity
\cite{1987ARA&A..25...23S,%
2001ApJ...562..852H}.

\subsection{Molecular clouds as part of a turbulent cascade}

Molecular clouds are known to be significantly supersonic,
with observed turbulent velocities of the order km\,s$^{-1}$,
while typical sound speeds at molecular cloud temperatures are
$\sim$ 0.2--0.3 km\,s$^{-1}$.  The supersonic velocities were
assumed to give rise to very rapid dissipation in shocks, and
hence explaining how the observed velocities are sustained
was regarded as a problem. A popular suggestion for a solution
to the problem was that the observed velocities are essentially
magneto-hydrodynamic waves, with very low dissipation rates
\cite{1975ApJ...196L..77A,1983ApJ...270..511Z}.

But in light of the conclusions of the previous section
there is really no reason to be concerned about how the
turbulence of molecular clouds is sustained; supply of kinetic
energy at molecular cloud size scales is available from the turbulent
cascade; i.e., simply from larger scale motions.  In fact, a
molecular cloud is probably borne precisely because the larger
scale velocity field happens to have a local maximum of
convergence there \cite{1999ApJ...515..286B,1999ApJ...527..285B,2001ApJ...562..852H}.

\subsection{Supersonic turbulent cascades}

\FIG{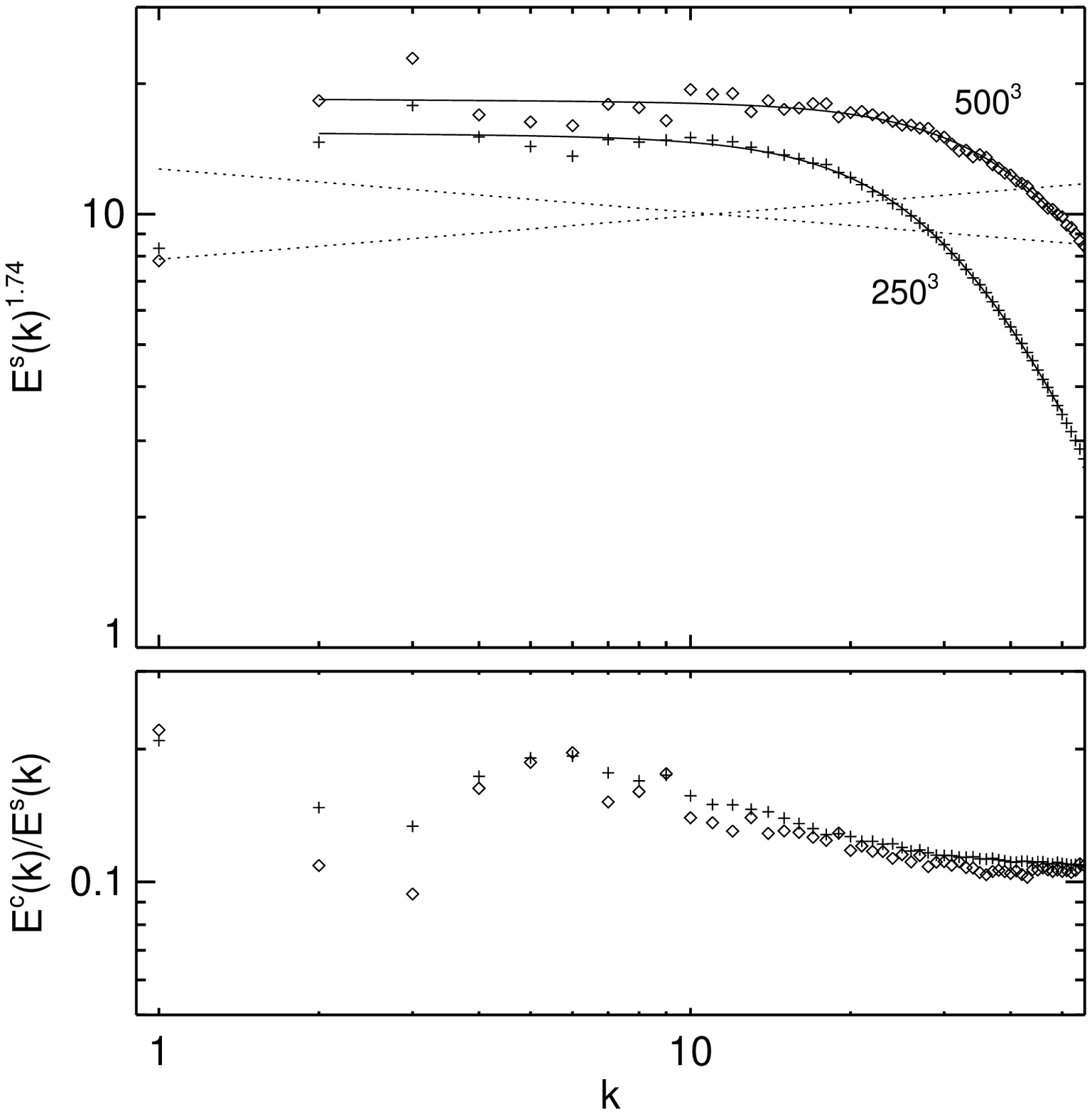}{0.8}{The solenoidal power spectrum, compensated by
$k^{1.74}$,  and the ratio of compressional to solenoidal power
in $\Mach \approx 10$ numerical experiments with
resolution $250^3$ (diamonds) and $500^3$ (stars), using
random solenoidal driving at { $1\leq k\leq2$}.  Dashed lines show
comparison slopes with spectral indices -1.64, -1.74, and
-1.84, respectively.}

\Figure{powersp} shows examples of power spectra of supersonic
turbulence, for two values of the numerical resolution.
The numerical simulations were performed with the same method as in
\cite{1997MNRAS.288..145P,1999ApJ...526..279P,Padoan+Corsica02,%
2001ApJ...553..227P}; a fifth order in space, third order in time
staggered mesh method, using per-unit-mass variables. The turbulence is
driven by a random external force that is applied in $1\leq k\leq2$ in
Fourier space. Only the solenoidal components of the force are used.
In order to ensure a smooth forcing the time derivatives of the Fourier
components of the force are regenerated at time intervals of about one
dynamical time and the force is computed from a time integral.

An inertial
(power law scaling) range is present, and extends to progressively
higher wavenumbers at higher numerical resolution.
How is it possible that supersonic turbulence gives rise
to a turbulent cascade much like that of incompressible
turbulence?  One clue comes from the dissipation rate of
supersonic turbulence.  Numerical experiments revealed it
to be similar to that of incompressible turbulence, if
expressed in terms of dynamical times $\tau_\textrm{dyn} =
\ell/v_\textrm{rms}(\ell)$, even for MHD-turbulence \cite{%
Biskamp+Muller00,%
MacLow_Puebla98,%
1998ApSS.261..195M,%
1999ApJ...524..169M,%
1999ApJ...526..279P,%
Stone+98}.
In qualitative terms one reason why the dissipation rate
is not as high as was naively expected is that shocks in
three-dimensional supersonic turbulence are typically
oblique rather than head-on, and that fragmentation decreases
the efficiency of interaction \cite{1982ApJ...258L..29S,%
1995MNRAS.277..377P}.

The ratio of compressional to solenoidal kinetic energy is small in
isotropic supersonic turbulence; typically $E_c/E_s\sim 0.1-0.2$
\cite{Boldyrev+1} (cf.\ \Fig{powersp}).
To appreciate why this is so it is helpful
to consider the velocities on either side of a shock sheet formed
by two interacting large scale streams.  By definition, the
gas upstream of the two stand-off shocks on either side of
the sheet have no casual connection, and their orientations
are therefore random with respect to each other.

\FIGG{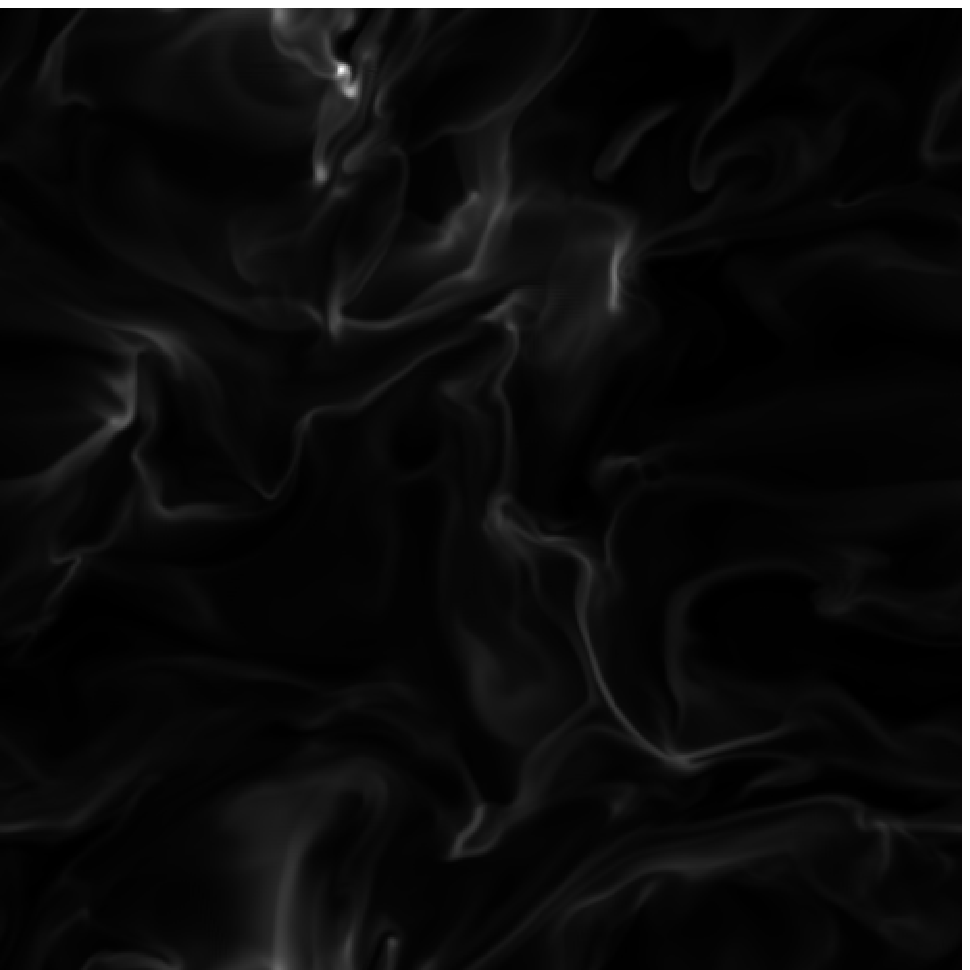}{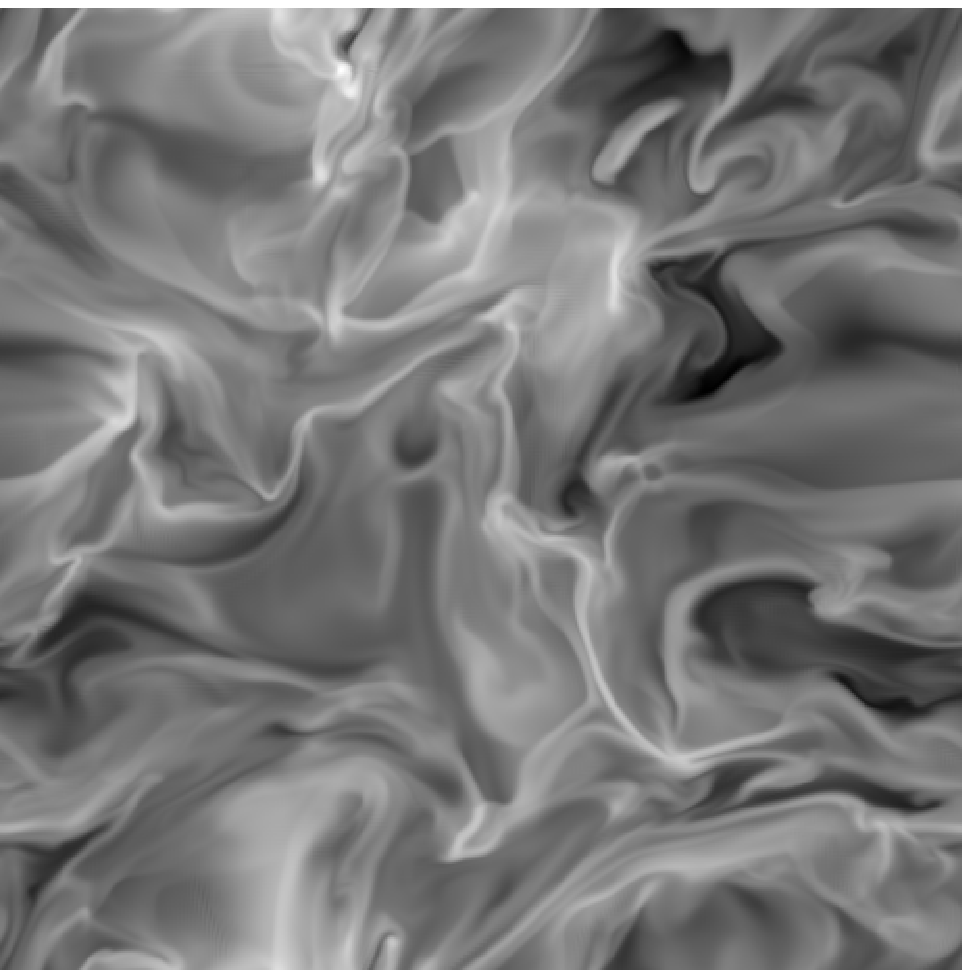}{.49}{Density in an arbitrary cut through
a numerical experiment with $\Mach \approx 10$ turbulence at a resolution
of 500x500x500.  a) Linear scaling (printed with gamma=0.7), normalized to the local
maximum of the density.  b) Logarithmic scaling, normalized to
the local maximum and minimum of the density.}

In terms of a coordinate system with one axis perpendicular to
the plane of the sheet, two of the three velocity components
are parallel to the sheet and hence incompressible, and only
one is perpendicular (compressional).  The compressional
component is the one that gives rise to the shock, with its
associated stagnant region inside the shock sheet.  It follows
that one should indeed expect the compressional component to,
on the average, carry less than one third of the kinetic
energy, consistent with what is found in the numerical
experiments.

More generally, one may think of three-dimensional supersonic
turbulence as an ensemble of shock sheets, and their associated
intersections in filaments and knots.  The typical history of
a trace particle in such a flow is that it participates in a
series of oblique shocks where, in each shock, the particle looses
some of its kinetic energy.

If the system consisted of an ensemble of stationary,
plane-parallel shock sheets a fluid parcel would first hit one
sheet, where its perpendicular kinetic energy would be
essentially lost.  It would then slide along the sheet until
it hit the filamentary intersection with another sheet, and
finally slide along the filament until it ended up in the
stagnant region of a knot-like intersection of filaments.

In a more general picture shock sheets are neither stationary
nor plane-parallel, which allows trace particles to participate
in a more extended series of shocks.  As its kinetic energy
is gradually reduced, so is the scale over which new shocks
are likely to be produced.

\Figure{cut-rho} illustrates the structure of the density field
in supersonic turbulence, modeled at a resolution of $500^3$.
The left hand side panel shows linear density.  Due to the
large density contrast, only a few shock sheets and filaments
are visible.  The right hand side panel shows logarithmic density,
and illustrates the general presence of intermittent density
structure over a range of scales and density levels.

\FIGGG{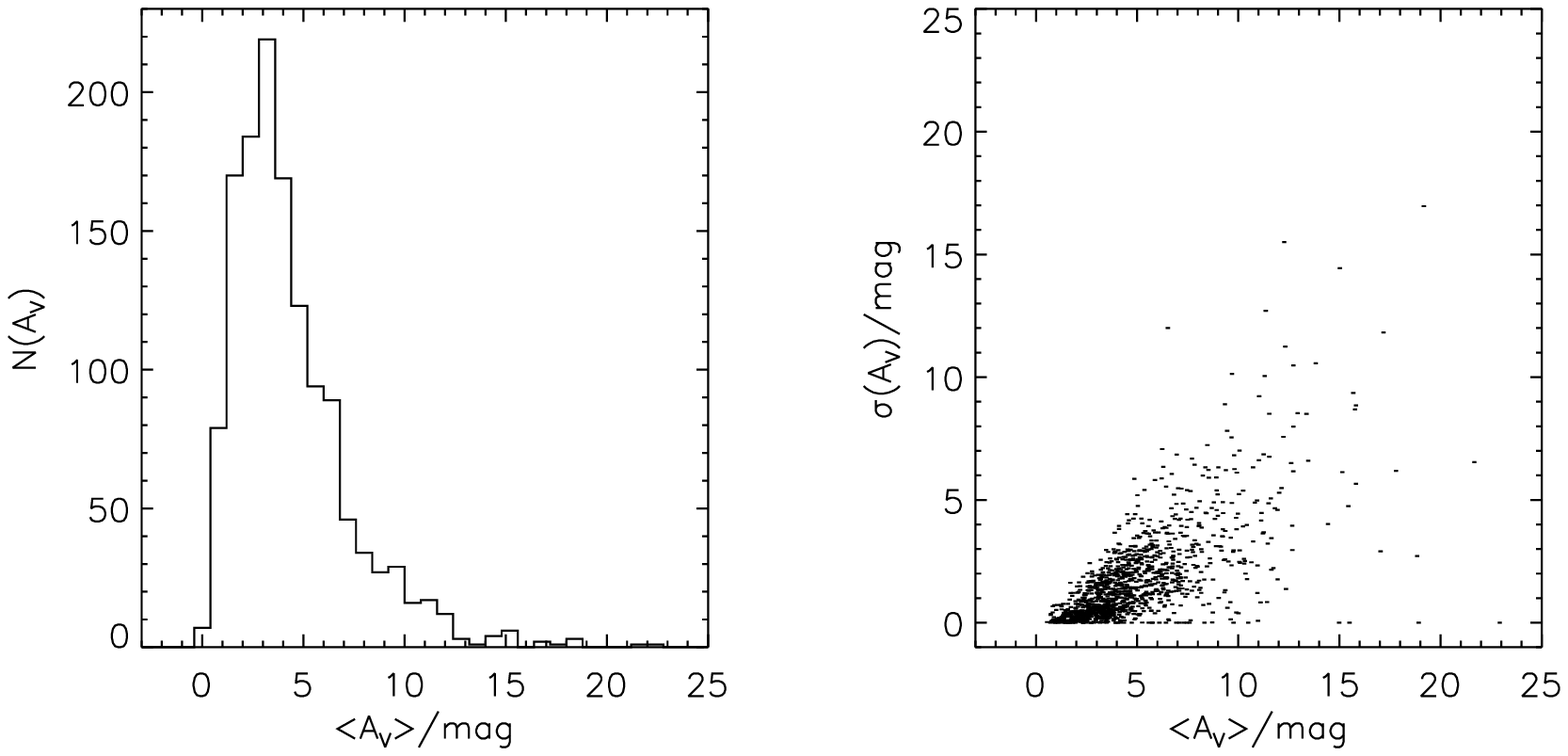}{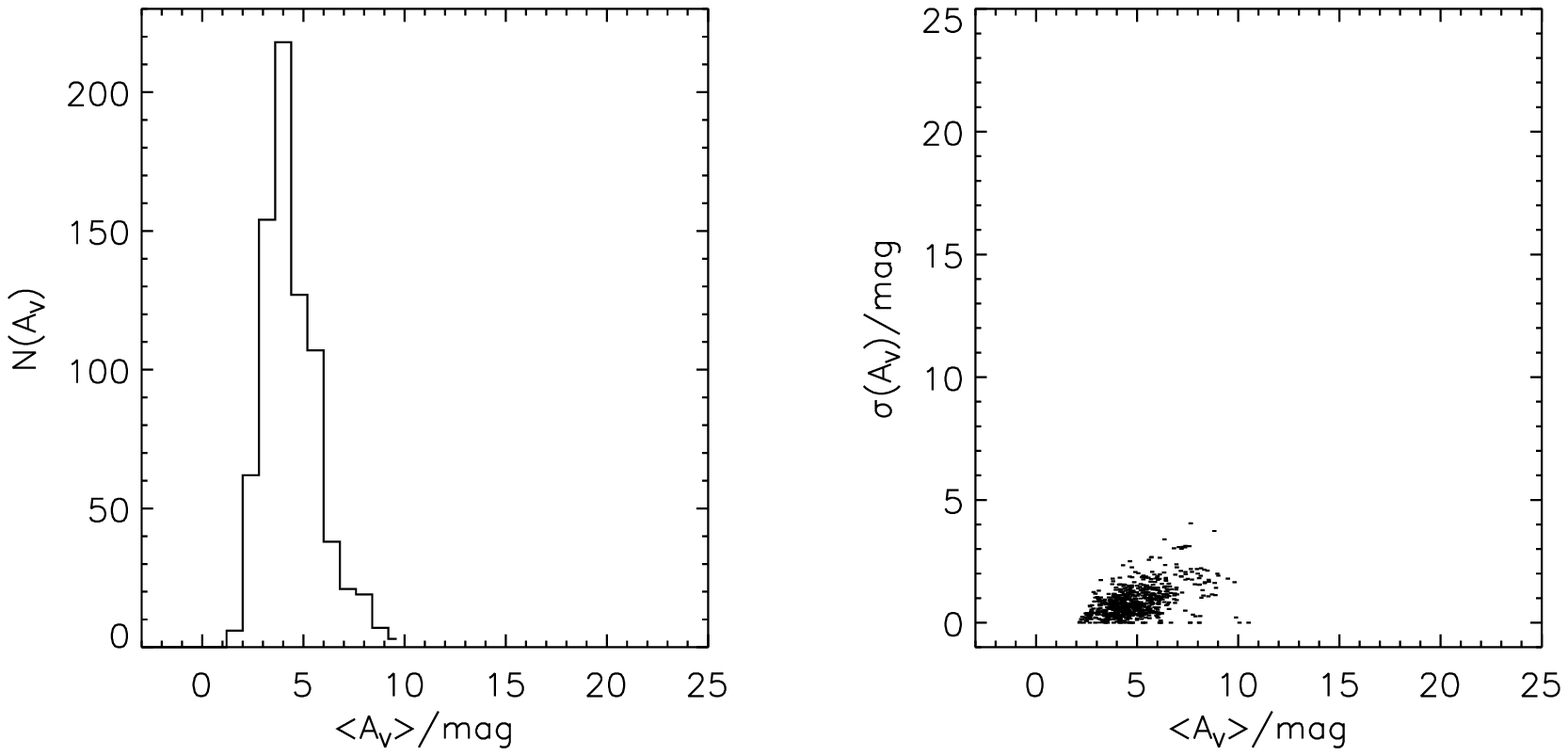}{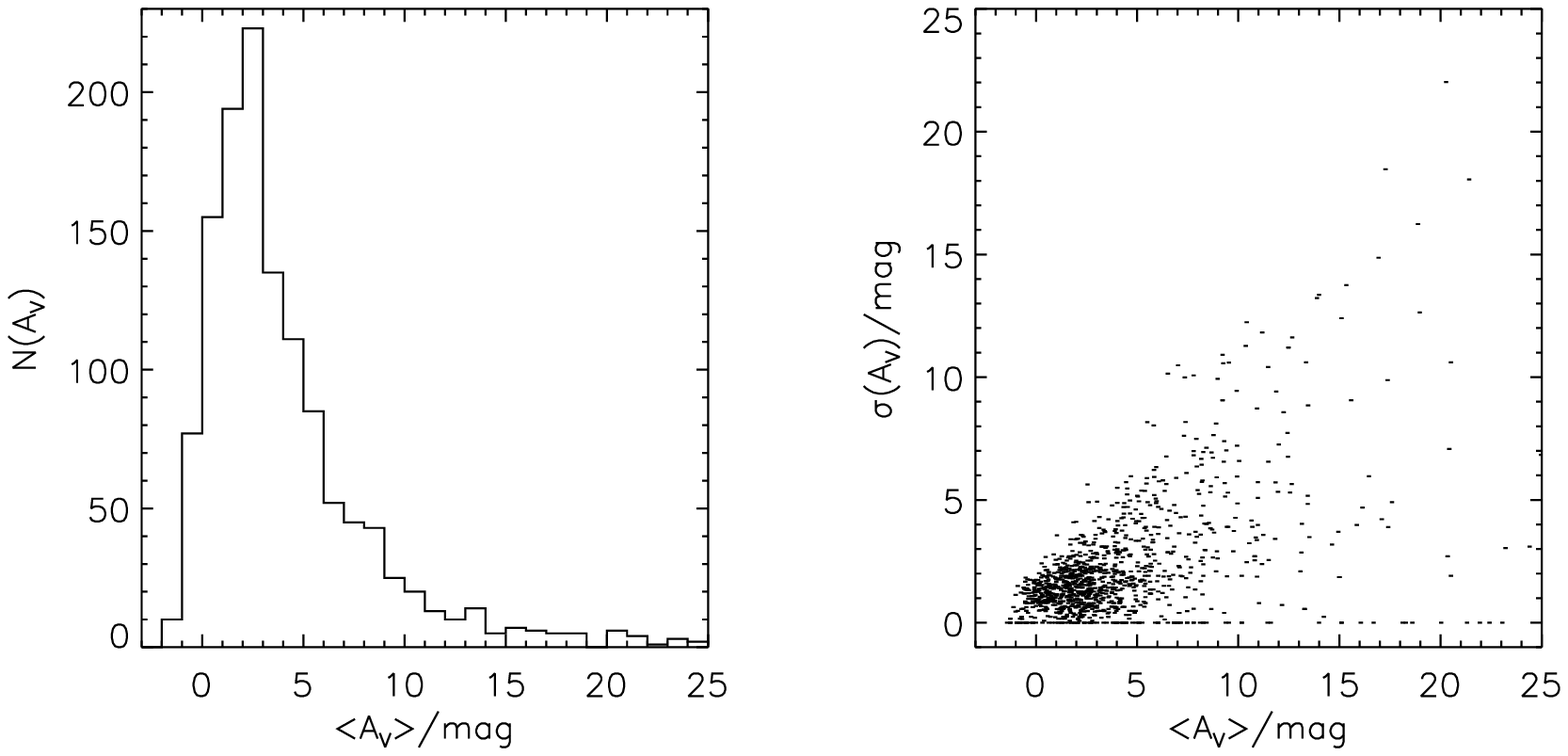}{0.44}
{Histograms of extinction (left panels) and plots of dispersion
of extinction in cells versus the mean cell extinction (right
panels). The top panels show the result from a
super--Alfv\'{e}nic model while the middle panels are from an
equipartition model \cite{1999ApJ...526..279P}. The bottom panels
are observational data for the cloud IC5146 \cite{Lada+94}}

\FIG{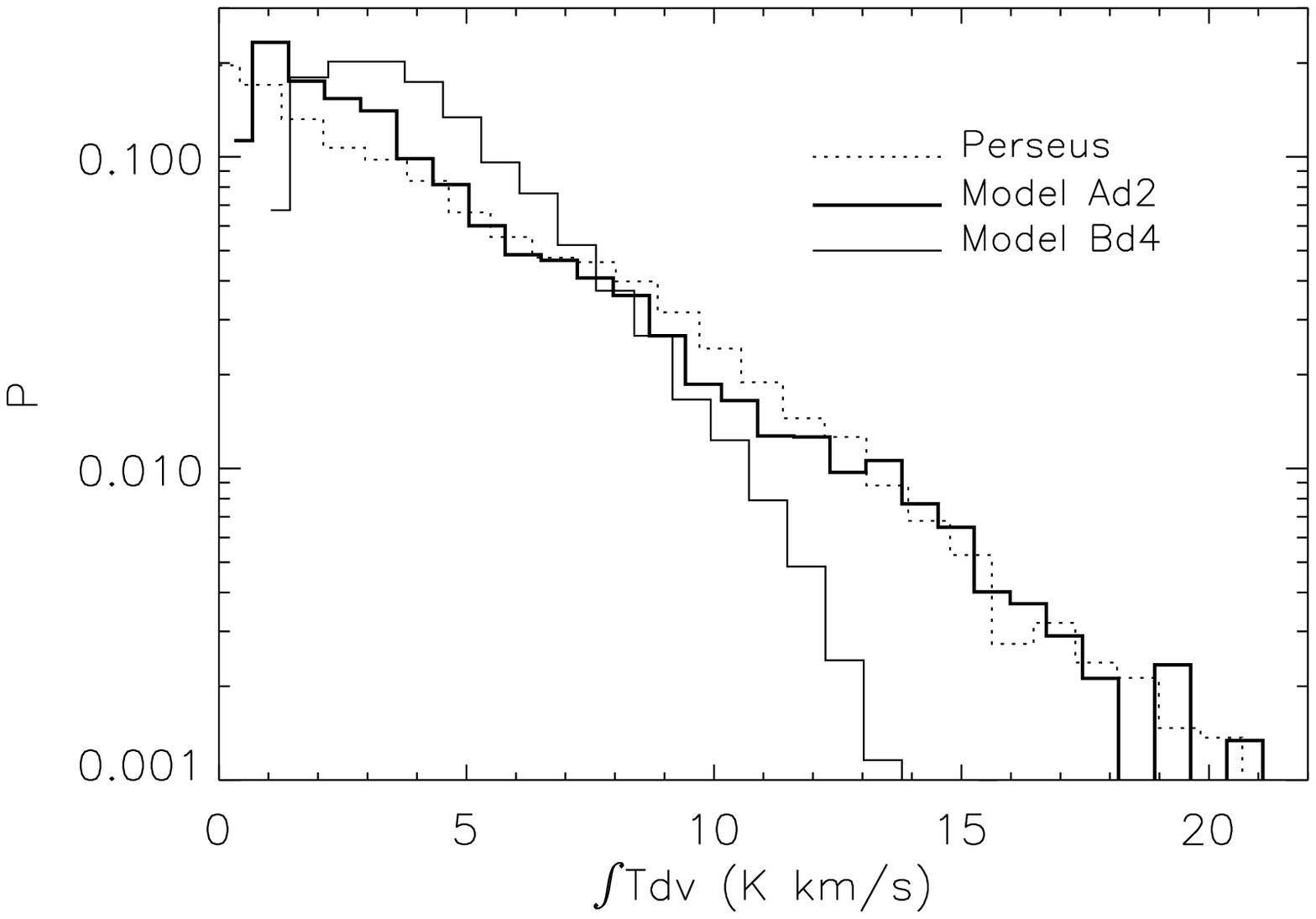}{0.6}{Histograms of integrated antenna temperature
of synthetic CO spectra \cite{1999ApJ...526..279P}, for a
super--Alfv\'{e}nic model (thick line), for an equipartition
model (thin line), and for the Perseus cloud (dotted line)}

The history of a fluid parcel in the real ISM might be
as follows:  It achieves initial, large kinetic energy by
either being part of the ejecta from a supernova or, more
likely, by being hit by the ejecta from a supernova.  It
becomes further compressed as the stream to which it belongs
collides with other streams.  As density increases cooling
becomes significant, and the temperature decreases.  The
parcel may eventually end up as part of a molecular cloud.
Inside the cloud, the process repeats itself, on successively
smaller scales, creating in the end a shock core massive enough
to form a star.  More likely, though, the fluid parcel
ends up in a structure too small to collapse by self-gravity,
where it survives until being hit by the blast wave from another
supernova, or the wind from a new-borne massive star.

\subsection{Super-Alfv{\'e}nic conditions}
\label{superA.sec}

The proposal that magneto-hydrodynamic waves are main
contributors to the velocity field in molecular clouds is now
obsolete for several independent reasons.  First, with the
velocity field of molecular clouds part of a turbulent
cascade there is no longer a problem to sustain the motions.
Second, with the demonstrations that MHD-turbulence decays
more or less as rapidly as hydrodynamic turbulence \cite{%
Biskamp+Muller00,%
MacLow_Puebla98,%
1998ApSS.261..195M,%
1999ApJ...524..169M,%
1999ApJ...526..279P,%
Stone+98}, %
the presumed `advantage' of MHD-turbulence has gone away.
Third, with observational and theoretical evidence that star
formation takes place on a time scale not much longer than a
crossing time \cite{2000ApJ...530..277E} the required life
times of molecular clouds are much shorter than was assumed in
earlier work.

\FIGGGG{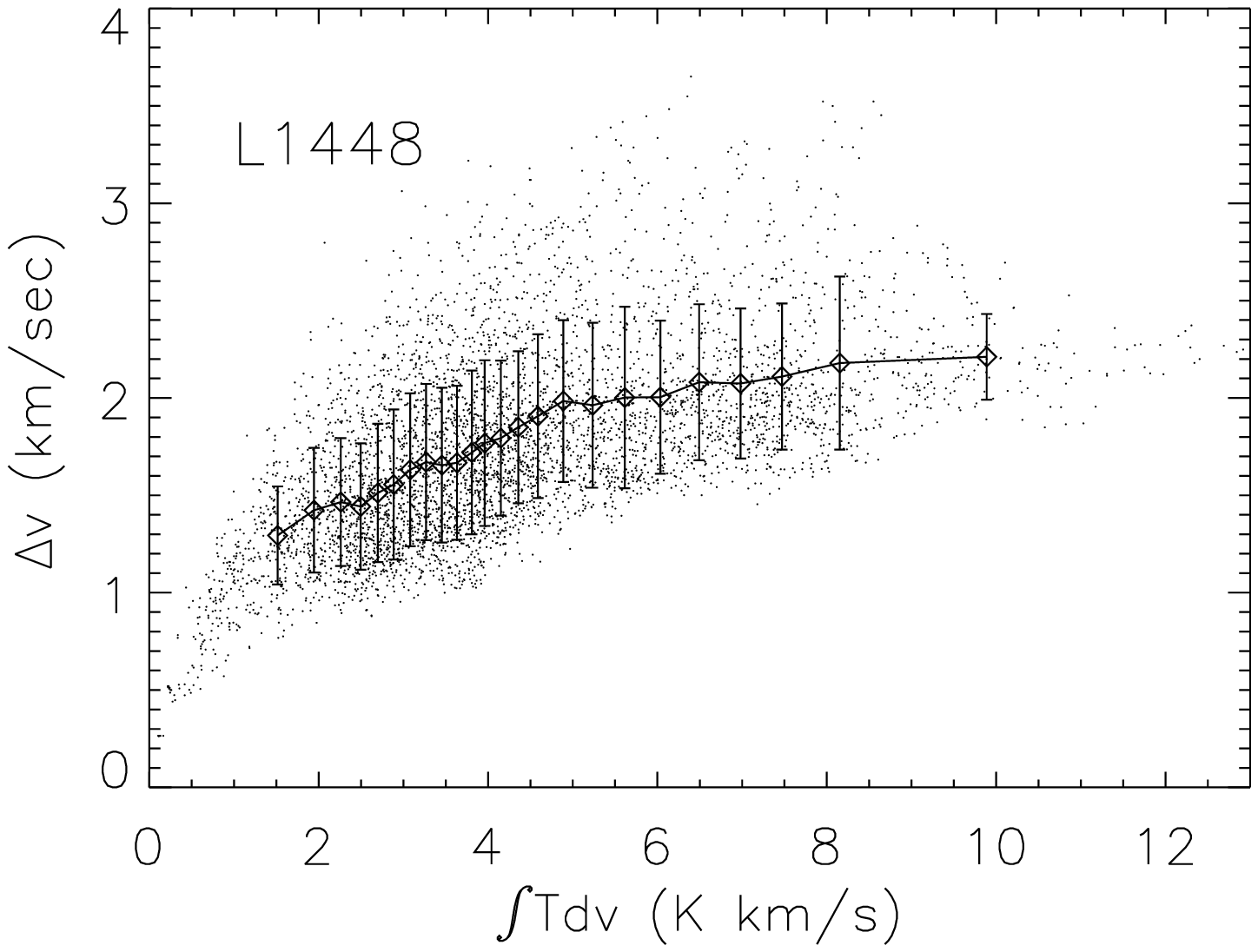}{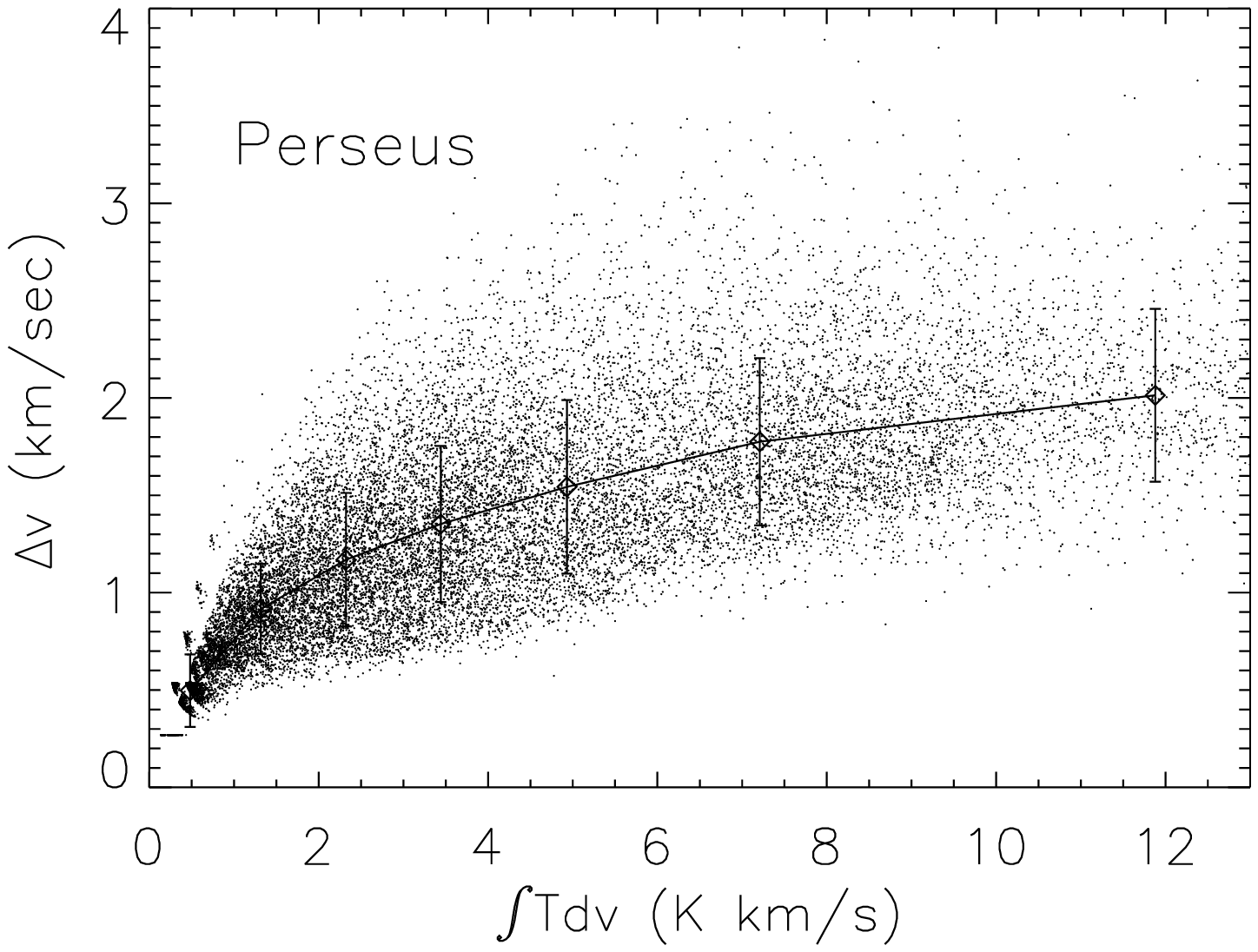}{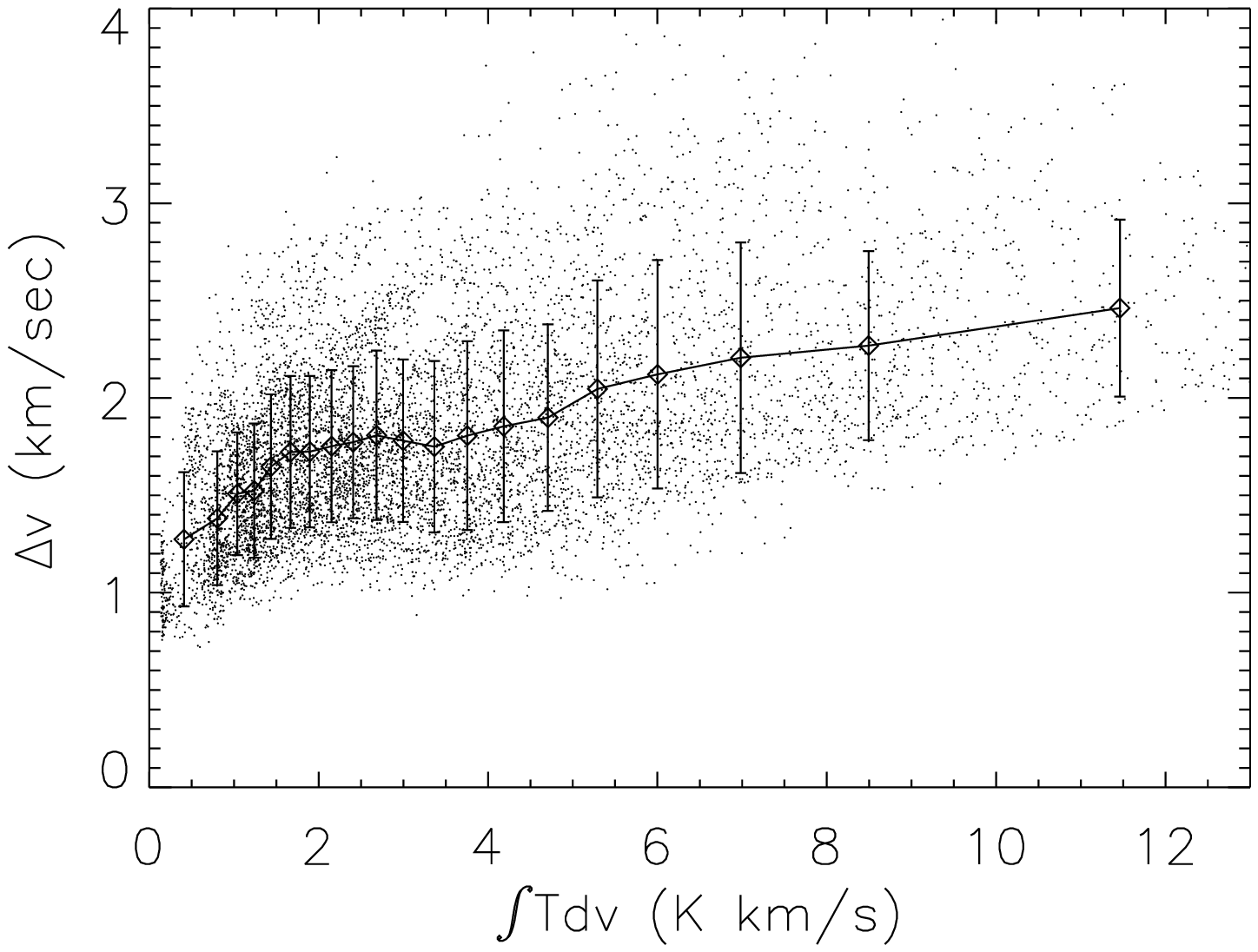}{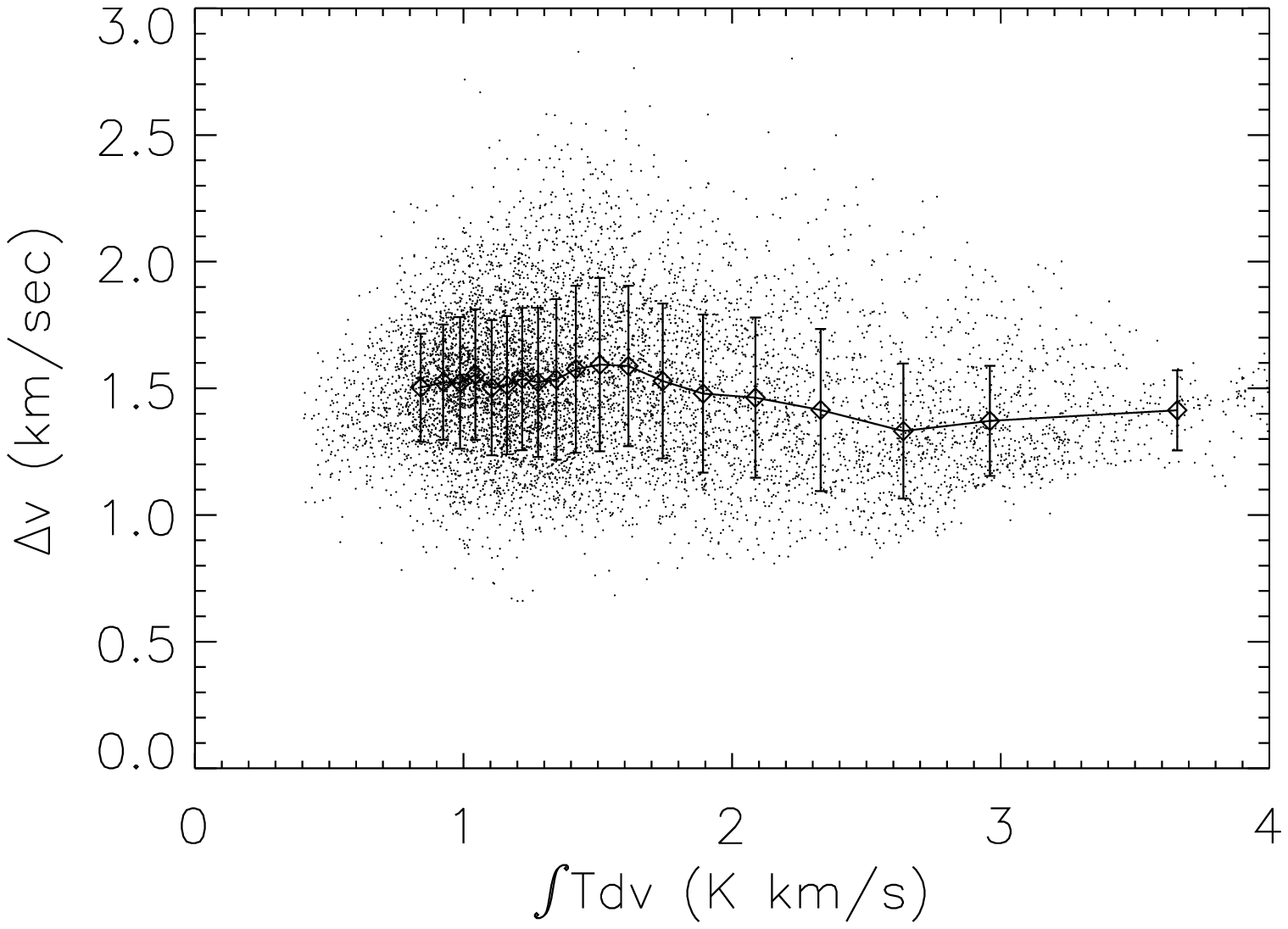}{0.35}
{Scatter plots of equivalent width versus velocity integrated
antenna temperature for two observed regions (two upper panels),
for a super-Alfvenic model, and for an equipartition mode.
The diamond symbols show the mean value of the equivalent width
in each interval of integrated antenna temperature and the
"error bars" show the one $\sigma$ distribution around the mean}

\FIG{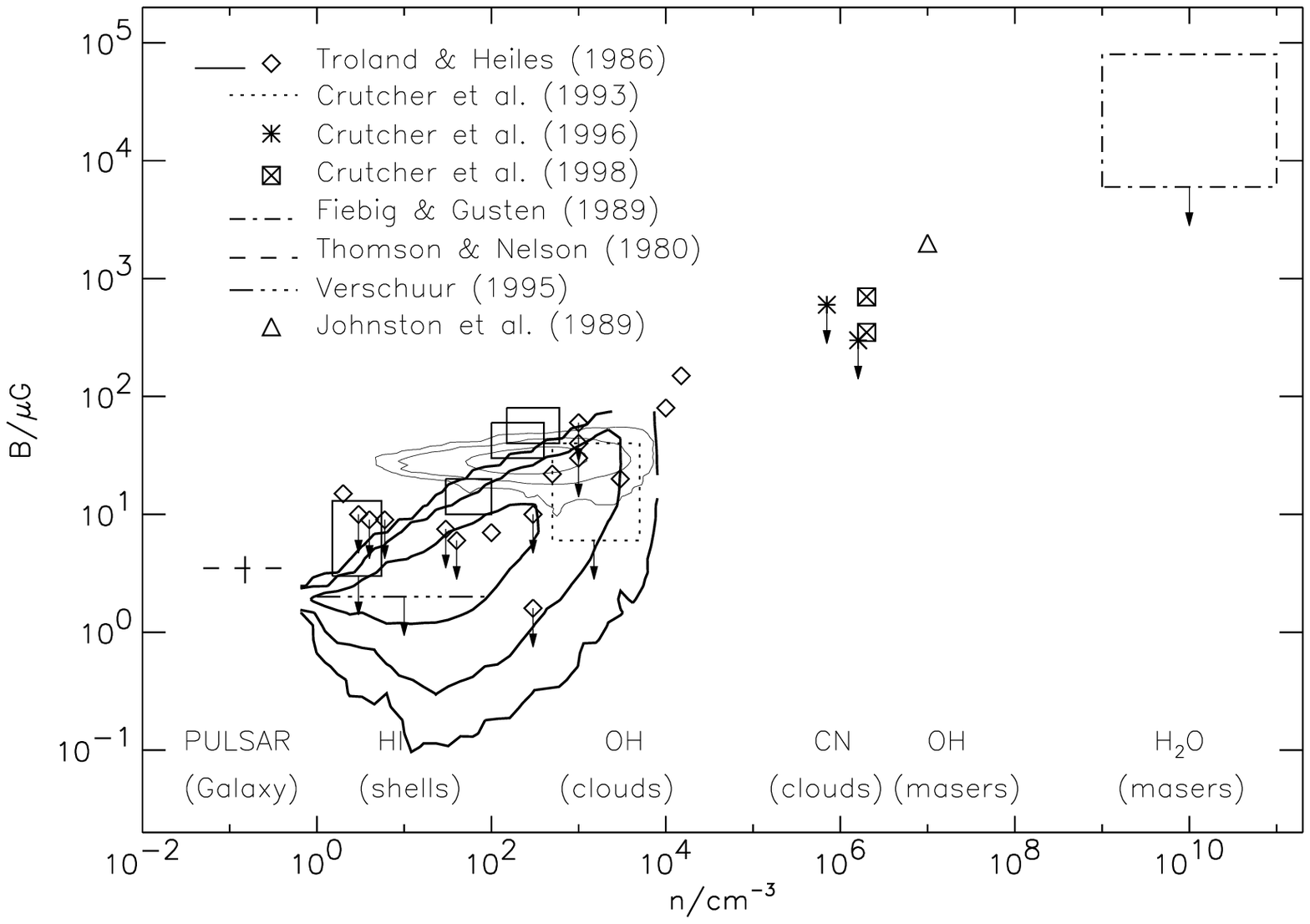}{0.6}
{The $B-n$ relation: observations and theoretical
models. The thick contour lines are from the a super--Alfv\'{e}nic
model and the thin contour lines are from an equipartition model
\cite{1999ApJ...526..279P}}


\FIG{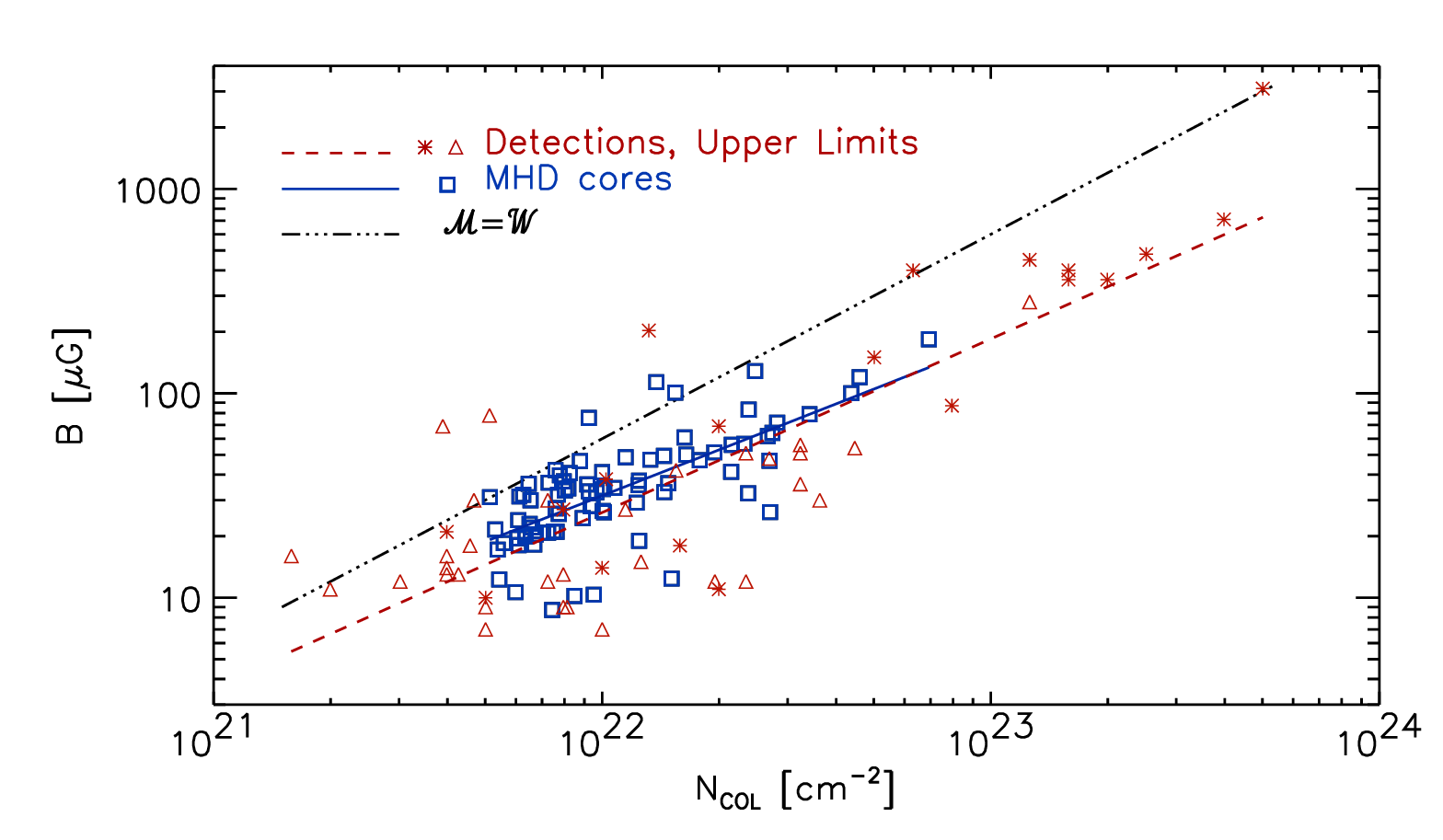}{0.51}{Magnetic field strength versus H$_2$ column
density.  Asterisks represent Zeeman splitting measurements
\cite{2001ApJ...554..916B} (detections and uppper limits)
with a least squares fit (dashed line).  Squares represent
cores from a super-Alfv{\'e}nic numerical experiment
(average B = 2.4 $\umu$G) \cite{Padoan+Corsica02},
with a least squares fit (solid line).
The dotted-dashed line marks equality between magnetic and
gravitational energies}

The notion that the velocity field in molecular clouds is
essentially Alfv{\'e}n waves lead to the assumption that the
kinetic and magnetic energy in molecular clouds are in near
equipartition.  Although inverting observations to obtain the
magnetic field strength, density and velocity in the same
structures is notoriously difficult (cf.\ the discussion in
Section 4.1 of \cite{Goodman+Heiles94}), equipartition
remained a popular null-hypothesis.

With access to numerical simulations it is possible to use the
more robust `forward analysis' method, where synthetic
diagnostics computed from the results of numerical simulations
are compared directly with the corresponding observational
diagnostics. Comparisons of extinction statistics, synthetic
molecular lines, the antenna temperature -- line width
relation, and the statistical upper envelope relation between
density and magnetic field strength (cf.\ \Figs{ext1}{Bn})
all lead to the same conclusion; models with
equipartition between magnetic and kinetic energy are
inconsistent with the observations while models where the kinetic
energy dominates over the magnetic energy (super-Alfv{\'e}nic
models) are consistent with the observations \cite{%
1997ApJ...474..730P,
1998ApJ...504..300P,
1999ApJ...525..318P,
1999ApJ...526..279P}
.

\FIG{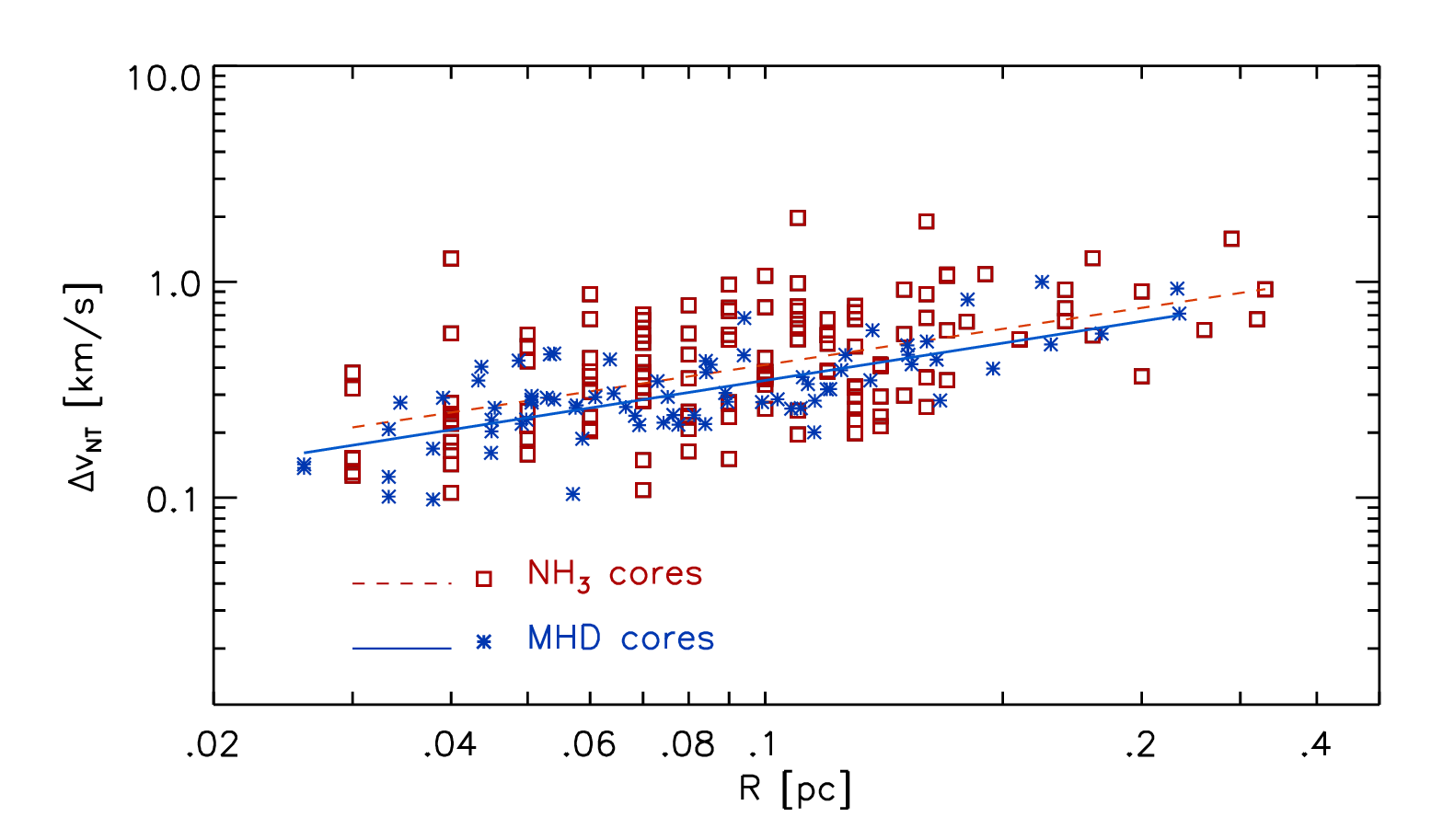}{.50}{Non-thermal line width of
observed (squares, with dashed line least squares fit)
and modeled (stars, with solid line least squares fit)
NH$_3$ cores versus their size \cite{%
1999ApJS..125..161J,%
Padoan+Corsica02}.
}

\FIG{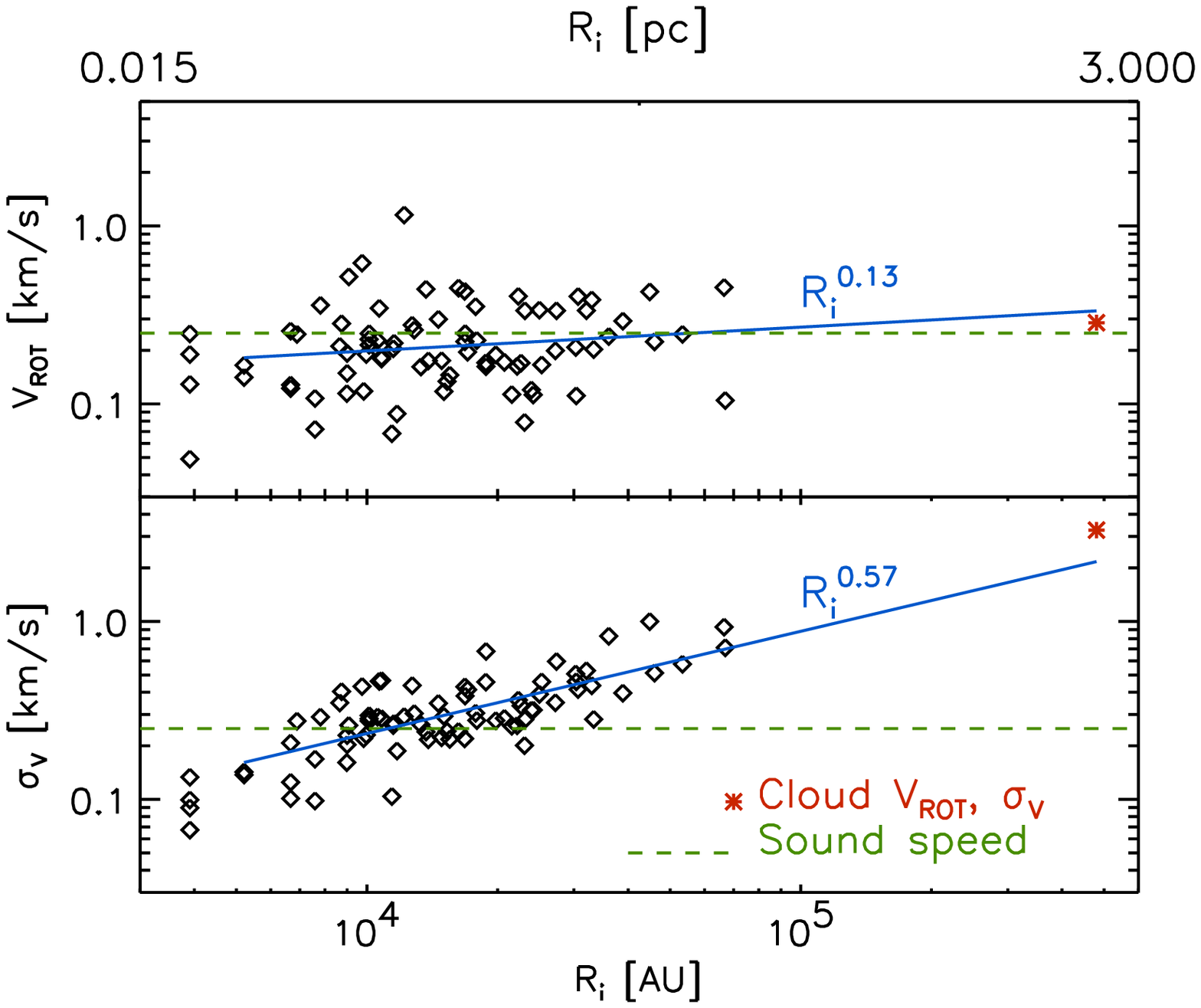}{.74}{Rotational velocity (upper panel) and
velocity dispersion (lower panel) for numerical cores.}

A direct illustration of the consistency of super-Alfv{\'e}nic
conditions with Zeeman observations of magnetic field strength
is given in \Fig{b-col}.  Note that several cores with B in
excess of 100 $\umu$G are found, even though the average B in
the simulation is only 2.4 $\umu$G.
This is a good example of the power of forward comparisons in
situations with strong intermittency; it would be very difficult to
recover the mean field strength, or the mean magnetic energy, directly
from the observations, which sample only the very small fraction of
the cloud volume filled by the densest regions. Further illustrations
are given in \Figs{larson-NH3}{larson-rot}, which show comparisons of
velocity statistics with observations.

\Figure{larson-NH3} is a comparison of the correlation of non-thermal line
width with size in NH$_3$ cores from the compilation by
Jijina, Myers \& Adams \cite{1999ApJS..125..161J} and in cores selected
from a simulation of supersonic and super-Alfv\'{e}nic turbulence
\cite{Padoan+Corsica02}. The least squares fit to the observational
data yields the power law exponent $0.56\pm 0.22$ and the fit to the
data the exponent $0.57\pm 0.15$. \Figure{larson-rot} shows the
correlation of rotational velocity (upper panel) and internal velocity
dispersion (lower panel) with size, for the numerical cores only.
The rotational velocities are very low and of the order of the sound
speed, as found in the observational data.
In both panels of \Fig{larson-rot} the asterisk symbols that correspond to
a size of almost 3~pc provide the values of rotational velocity and
velocity dispersion computed over the whole simulated volume.
Although the least squares fits are computed only for the cores,
the values for the whole system are consistent with the fits.

\subsection{The magnetic flux problem}

A ``magnetic flux problem'' is often mentioned in this context.  It
is argued that since the average density of molecular clouds is at
least 100 times larger than the average density in the galactic disk,
and assuming magnetic flux freezing, the average magnetic field strength
of molecular clouds should be much larger than the average galactic
values of a few $\umu$G.  As demonstrated by \Fig{b-col} there is in
fact no real problem -- observations of core magnetic fields are
completely consistent with predictions from models with average
magnetic field strengths of a few $\umu$G -- there is at most a
conceptual / perceived problem.

This conceptual problem has a straightforward solution, which,
ironically, is most easily demonstrated by the equipartition model. It
has been shown in many numerical works that supersonic turbulence, even
with equipartition of kinetic and magnetic energy (the traditional
model for molecular clouds) generates a complex density field, with
very large contrast sheetlike and filamentary density structures.
These density enhancements do \emph{not} correspond to significant
variations of the magnetic field strength in equipartition models,
since in them strong compression can occur only along magnetic field
lines. To the extent that turbulence on large scales (disk thickness)
has approximate equipartition of kinetic and magnetic energy, molecular
clouds can still easily form, as a consequence of compressions
\emph{along} magnetic field lines.

Once a cloud is formed by large scale equipartition turbulence, it has
a mean magnetic field strength close to the galactic value, and its
internal dynamics is super-Alfv\'{e}nic, because of the much increased
density.  Equipartition on the large scale, therefore, is not a problem
for the origin of super-Alfv\'{e}nic clouds.

The argument applies recursively; should the super-Alfv\'{e}nic cloud
by chance create a region with local equipartition, further increase of
the density is still possible, by inflow of mass along magnetic field
lines.  One sees the statistical outcome of this in the B-n relation
(\Fig{Bn}); for any given density there is a wide distribution of
magnetic field strengths, up to an upper envelope given by approximate
equipartition.

Inflow along magnetic field lines is also likely to occur in the
phase when gravitation has taken over, after local cores are formed
along filaments and in corrugated sheets.  In that situation the
magnetic field has already been compressed, and is oriented
predominantly along the same filaments and sheets that also
contain abundant mass at high density.

\subsection{Gravitationally bound and unbound clouds}

The scenario where star formation takes place in essentially
a crossing time \cite{2000ApJ...530..277E} also alleviates
earlier concerns about how to support molecular clouds
against gravitational collapse. The gravitational binding
energy of molecular clouds is often comparable to their turbulent
kinetic energy, and hence quite a bit larger than their thermal
energy \cite{1981MNRAS.194..809L,1992A&A...257..715F}. This raised
the question of the support of the clouds against gravitational
collapse.  Could the turbulent velocities be translated into a
turbulent pressure that was able to support the clouds against collapse
\cite{%
1987A&A...172..293B,%
1992JFM...245....1B,%
1995A&A...303..204V,%
2000ApJ...535..887K}?
Or did the
solution lie in the observed, strong fragmentation of the medium
\cite{1982ApJ...258L..29S,1995MNRAS.277..377P}?

In the `turbulent fragmentation \& star formation in a
crossing time' paradigm it is natural to find some molecular
clouds with roughly virial mass, as well as some with
substantially less than virial mass, while there are
essentially none with much larger than virial mass (cf.\ Fig.\ 7
of \cite{1992A&A...257..715F}).  Turbulent fragmentation
creates clouds, initially without regard to gravity.  Some
of the clouds that are produced are gravitationally unbound
(but may contain sub-structures that are gravitationally
bound).  Some other clouds are massive enough to be gravitationally
bound (at least until their first supernovae blow out
a major part of the cloud gas).  Clouds that are created with
a mass larger than virial start to collapse, which increases
the velocity dispersion of their sub-structures until they
appear to be essentially virial.

The latter case represents the most direct and simple
mechanism by which turbulence prevents global collapse of
molecular clouds; i.e., through fragmentation rather than through
``turbulent pressure''.  The mechanism may be illustrated by
considering that extreme intermittency caused by strongly
supersonic turbulence and cooling could create conditions where
individual density maxima move in essentially ballistic orbits
relative to one another \cite{1982ApJ...258L..29S}.  Even under
less extreme conditions intermittency may cause individual
density maxima to collapse, while the cloud as such does not
\cite{1995MNRAS.277..377P,%
1997MNRAS.288..145P,%
2000ApJ...535..887K,%
2001ApJ...547..280H}.

\subsection{Power laws and equipartition}

As mentioned above, even highly supersonic turbulence is
characterized by power laws \cite{Boldyrev+1,Boldyrev+2}.
However, because of the strong intermittency of density and its
correlation with the velocity field, the spectrum of kinetic
energy is not the same as the power spectrum of velocity.

It is appropriate to define the spectrum of kinetic energy
as the power spectrum of $\rho^{1/2} \vec{u}$, since the
sum of squares of its Fourier components is equal to
the kinetic energy.  Empirically, from numerical simulations,
one finds that the spectrum
of kinetic energy is quite a bit more shallow than the power
spectrum of velocity.  The latter has a power exponent consistent
with the theoretical
expectation $\beta = 1.74$ \cite{Boldyrev+1}.   The former has
a power exponent $\beta_{\mathrm{k.e.}} \approx 1.1$.

The power spectrum of the magnetic field is approximately parallel
to that of velocity in the high-k part of the inertial range, and hence the spectrum
of magnetic energy is \emph{steeper} than the spectrum of kinetic
energy.  This may appear strange, at first.  Why would the magnetic
field have a power spectrum similar to that of velocity, when magnetic
energy, $\half B^2$, is measured in the same units as kinetic energy
$\half\rho u^2$ and not in the units of velocity power $u^2$?
A possible explanation is that $B^2$ is weighted more towards the
kinetic energy of the bulk of the volume.  Assuming a log-normal
PDF of density, with a dispersion of linear density \cite{1999intu.conf..218N}
\EQ{sigma-dens}
\sigma_\rho = b\Mach ~,
\EN
where $b \approx \half$,
the most common density is
\EQ{rho-0}
\rho_\mathrm{0} = {\langle\rho\rangle \over (1+b^2\Mach^2)^\half} ~,
\EN
which is smaller than the average density $\langle\rho\rangle$.
Thus, if the magnetic energy is in equipartition with the kinetic
energy at those densities, rather than at the higher densities
towards which the average kinetic energy is weighted, this would
explain both why the spectrum of $\vec{B}$ is similar to that
of $\vec{u}$ and why the average magnetic energy is below
equipartition with the kinetic energy.

With a difference in the power law exponents a gap develops from
the (observed) equipartition at large scales ($\sim$ 100 pc).  A
power law index difference of 0.65 implies $E_{mag}/E_{kin} < 0.1$
at pc scales, which is consistent with the forward analysis of
numerical simulations \cite{1999ApJ...526..279P}.

Note that the discussion above is complementary to the one at the
end of \Sect{superA} -- both views are helpful for understanding
why small scale ISM motions are super-Alfv{\'e}nic.

\section{A new analytical theory of supersonic turbulence}
\label{stas.sec}

Due to the complexity of the Navier-Stokes equations, mathematical work
on turbulence is often inspired by experimental and observational
measurements. Since geophysical and laboratory flows are predominantly
incompressible, turbulence studies have been limited almost entirely
to incompressible flows (or to infinitely compressible ones, described
by the Burgers equation). Little attention has been paid in the past to
highly compressible, or super-sonic turbulence.

Turbulent flows are traditionally described statistically by the
structure functions of their velocity field \cite{Frisch95}.
The structure functions are defined as
\begin{equation}
S_p(\ell)=\langle |\vec{u}(\vec{x}+\vec{\ell})-\vec{u}(\vec{x}))|^p\rangle \propto L^{\zeta(p)}
~,
\label{structure}
\end{equation}
where $\vec{u}$ is the component of the velocity field perpendicular
(transversal structure functions) or parallel (longitudinal
structure functions) to the vector $\vec{\ell}$. In the inertial
interval the structure functions obey scaling laws and the
exponent $\zeta(p)$ can be determined. The power spectrum
of the velocity is the Fourier transform of the second order
structure function, and may be expressed as
$E(k)\propto k^{-\beta}\propto k^{-1-\zeta(2)}$.

One may think that the study of high order structure functions
is interesting only for testing models of intermittency in turbulent
flows, and not very useful in the context of ISM turbulence and star
formation. Actually, the intermittent nature of turbulence is crucial
in modeling the process of star formation driven by turbulent fragmentation.
Stars are formed in the densest regions of turbulent flows.
These regions contain only a few percent of the total mass and fill
an almost insignificant fraction of the total volume of a star forming
cloud. High order moments defining the tails of
statistical distributions of velocity and density are therefore very
important in the process of star formation. Furthermore, low order
density structure functions, which are obviously important to describe
basic properties of turbulent fragmentation, can be shown to depend on
velocity structure functions of very high order \cite{Boldyrev+2}.

The scaling of the velocity structure functions in incompressible
turbulence is best described by the She-Leveque formula \cite{SL94},
\begin{equation}
\frac{\zeta(p)}{\zeta(3)}=p/9 + 2\left[ 1-\left( \frac{2}{3}\right) ^{p/3}\right]
~.
\label{s-l}
\end{equation}
The scaling exponents are computed relative to the third order,
$\zeta(p)/\zeta(3)$, because according to the
concept of extended self-similarity \cite{Benzi+93,Dubrulle94}
the relative exponents are universal and better defined than the
absolute ones.

Boldyrev \cite{Boldyrev} has proposed an extension of the She-Leveque's
formalism \cite{SL94} to the case of supersonic turbulence. Based
on the physical interpretation of (\ref{s-l}) by Dubrulle \cite{Dubrulle94},
a fundamental parameter in the derivation of the velocity structure
functions is the Hausdorff dimension of the support of the most
singular dissipative structures in the turbulent flow. In
incompressible turbulence the most dissipative structures are
organized in filaments along coherent vortex tubes with Hausdorff
dimension $D=1$, while in supersonic turbulence dissipation
occurs predominantly in sheet-like shocks, with Hausdorff dimension
$D=2$. The new velocity structure function scaling proposed by Boldyrev
\cite{Boldyrev} for supersonic turbulence is
\begin{equation}
\frac{\zeta(p)}{\zeta(3)}=p/9 + 1-\left( \frac{1}{3}\right) ^{p/3} ~.
\label{boldyrev}
\end{equation}
This velocity scaling has been found to provide a very accurate
prediction for numerical simulations of supersonic and
super-Alfv\'{e}nic turbulence \cite{Boldyrev+1}, and has been
used to infer the structure of the density distribution
in turbulent clouds \cite{Boldyrev+2}.

\section{Star formation and the Initial Mass Function}
\label{imf.sec}

At least three unrelated ways of explaining the process of star formation
and the origin of the stellar initial mass function (IMF) may be
found in the literature: i) Ambipolar drift contraction of
sub-critical cores \cite{%
1987ARA&A..25...23S,
1996ApJ...464..256A}; 
ii) opacity-limited gravitational fragmentation
\cite{1953ApJ...118..513H,%
1963ApJ...138.1050G,%
1972PASJ...24...87Y,%
1976SvA....19..403S,%
1976MNRAS.176..367L,%
1976MNRAS.176..483R,%
1977ApJ...214..718S,%
1977ApJ...214..152S,%
1985ApJ...295..521Y};
and iii) turbulent fragmentation
\cite{1971ApJ...169..289A,%
1981MNRAS.194..809L,%
1982ApJ...258L..29S,%
1993ApJ...419L..29E,%
1994ApJ...423..681V,%
1995MNRAS.277..377P,%
2001ApJ...553..227P,%
Padoan+Nordlund-IMF}.

The first type of models rely on the assumption that both protostellar
cores and their parent clouds are long lived systems in near equilibrium,
supported against their gravitational collapse by magnetic field pressure.
As discussed above, this assumption has been proven incorrect based on
observational data and is inconsistent with the turbulent nature
of star-forming clouds \cite{%
1999ApJ...526..279P,%
2000ApJ...530..277E,%
2001ApJ...562..852H,%
2001ApJ...553..227P}.
Furthermore, these type of models
do not address the problem of the formation of massive stars or brown
dwarfs, and have traditionally focused more on the evolution of
individual protostars, without providing a self-consistent picture for
the origin of the initial conditions.

The second type of models is also inconsistent with the properties of
star-forming clouds, because it applies the concept of gravitational
fragmentation to the large scale, in the attempt of modeling the formation
of a whole stellar population. The concept of gravitational instability
is based on a comparison between the gas thermal and gravitational energies
to define the smallest unstable mass, or Jeans' mass. However, star-forming
clouds, as any region of the cold ISM above a scale of approximately 0.1~pc,
contain a kinetic energy of turbulence that is much larger (typically 100
times larger)
than their thermal energy, making the comparison of thermal and gravitational
energies irrelevant on the large scale. Attempts to redefine the Jeans' mass
\cite{Chandrasekhar58,1971ApJ...169..289A}
assuming that turbulence can provide pressure support against the
gravitational collapse are flawed, because they miss the basic point that
supersonic turbulence is actually fragmenting the gas. The main effect
of the large kinetic energy of turbulence, relative to the thermal
energy, is that the gas density and velocity fields in star-forming
regions are highly non-linear, against the assumption of the gravitational
instability model. In other words, clouds are already fragmented by
turbulence, quite independent of their self-gravity.

The third type of models, which we refer to as {\it turbulent fragmentation}
models, focus on the importance of the observed supersonic turbulence
in molecular clouds and are therefore consistent with the large scale
dynamics of star-forming regions. The idea of star formation
driven by supersonic turbulence was proposed twenty
years ago by Larson \cite{1981MNRAS.194..809L}, but has become popular
only in the last few years, thanks to the progress of numerical simulations
of supersonic magneto-hydrodynamic (MHD) turbulence.

According to the model of turbulent fragmentation, protostellar
cores are formed from gas compressed by shocks in the
supersonic turbulent flow \cite{2001ApJ...553..227P}.
While scale-free turbulence generates a power law mass
distribution down to very small masses, only cores with a gravitational
binding energy in excess of their magnetic and thermal energy can collapse.
The shape of the stellar IMF is then a power law for large masses, since
the majority of large cores are larger than their Jeans' mass.
At smaller masses, the IMF flattens and then turns around according
to the probability of small cores to be dense enough to collapse,
which is determined by the PDF of gas density.

\subsection{The Initial Mass Function}

The mass distribution of dense cores formed in a supersonic turbulent
flow follows from a number of properties of such flows
\cite{Padoan+Nordlund-IMF}: i) The power spectrum of the turbulence is
a power law; ii) the dynamics on scales covered by the power law is
approximately selfsimilar; iii) the typical size of a dense core scales
as the thickness of the postshock gas; iv) the relevant shock jump
conditions are those of MHD shocks.

These properties are to a considerable extent already verified by numerical simulations,
which also produce corresponding numerical IMFs \cite{Padoan+Corsica02},
but for the purpose of deriving a theoretical IMF one may just adopt
them as assumptions \cite{Padoan+Nordlund-IMF}.

The second assumption, about approximate selfsimilarity, is a crucial one.
Since velocity amplitudes do depend on scale (by the first assumption
and by Larson's relation) it is not an exact property.  Nevertheless,
since these flows are supersonic they follow essentially inertial
paths in a large fraction of space (upstream of shocks).  The thickness
and density of the downstream, shocked gas does depend on the Mach
number, and hence on the scale, but the filling factor of the shocked
gas is quite small and does not disturb the overall selfsimilarity much.

The distribution of cores that form in the shocked, downstream gas may
be regarded as a distribution over linear sizes $L$ of the upstream
flows out of which they formed.  By the assumption of approximate
selfsimilarity the number of such regions per unit log $L$ scales as
$L^{-3}$. The upstream (Alfv{\'e}nic) Mach number is denoted $\Mach(L)$
and is assumed to scale as $L^{\alpha}$, where $\alpha$ is related to
the power spectrum index $\beta=1+2\alpha$.

The typical mass of the cores that form in the shocked gas scales as
$\lambda^3 \rho_1$, where (from the MHD shock jump conditions) $\lambda
\sim L/\Mach(L)$ is the thickness of the postshock gas, $\rho_1 \sim
\rho_0 \Mach(L)$ is its density, where $\rho_0$ is the upstream mass
density (similar to the mean density). The typical core mass is thus
\begin{equation}
   m(L) \sim \rho_0 L^3/\Mach(L)^2 \sim L^{3-2\alpha} ~,
\end{equation}
which leads to the following
expression for the mass distribution of dense cores:
\begin{equation}
   N(m) \sim L(m)^3 \sim m^{-3/(3-2\alpha)} \sim m^{-3/(4-\beta)} ~.
\label{imf}
\end{equation}
If the power spectral index $\beta$ is consistent with the
observed velocity dispersion-size Larson relation \cite{1981MNRAS.194..809L}
and with the numerical and analytical results \cite{Boldyrev,Boldyrev+1},
then $\beta \approx 1.74$ and the mass distribution is
\begin{equation}
N(m)\,{\rm d}\log m\propto m^{-1.33}{\rm d}\log m ~,
\label{salpeter}
\end{equation}
which is almost identical to the Salpeter stellar IMF
\cite{1955ApJ...121..161S}.

\FIGGG{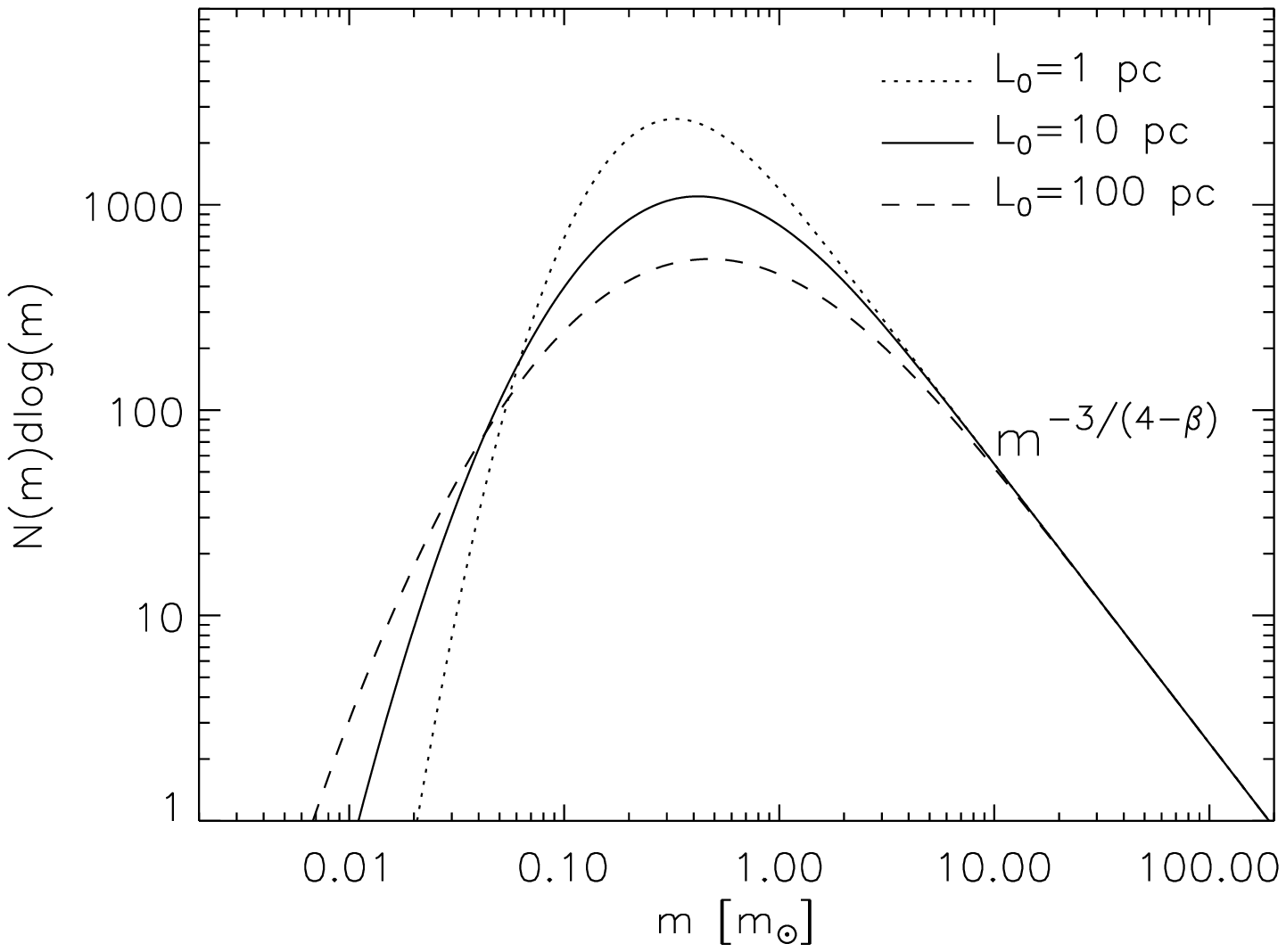}{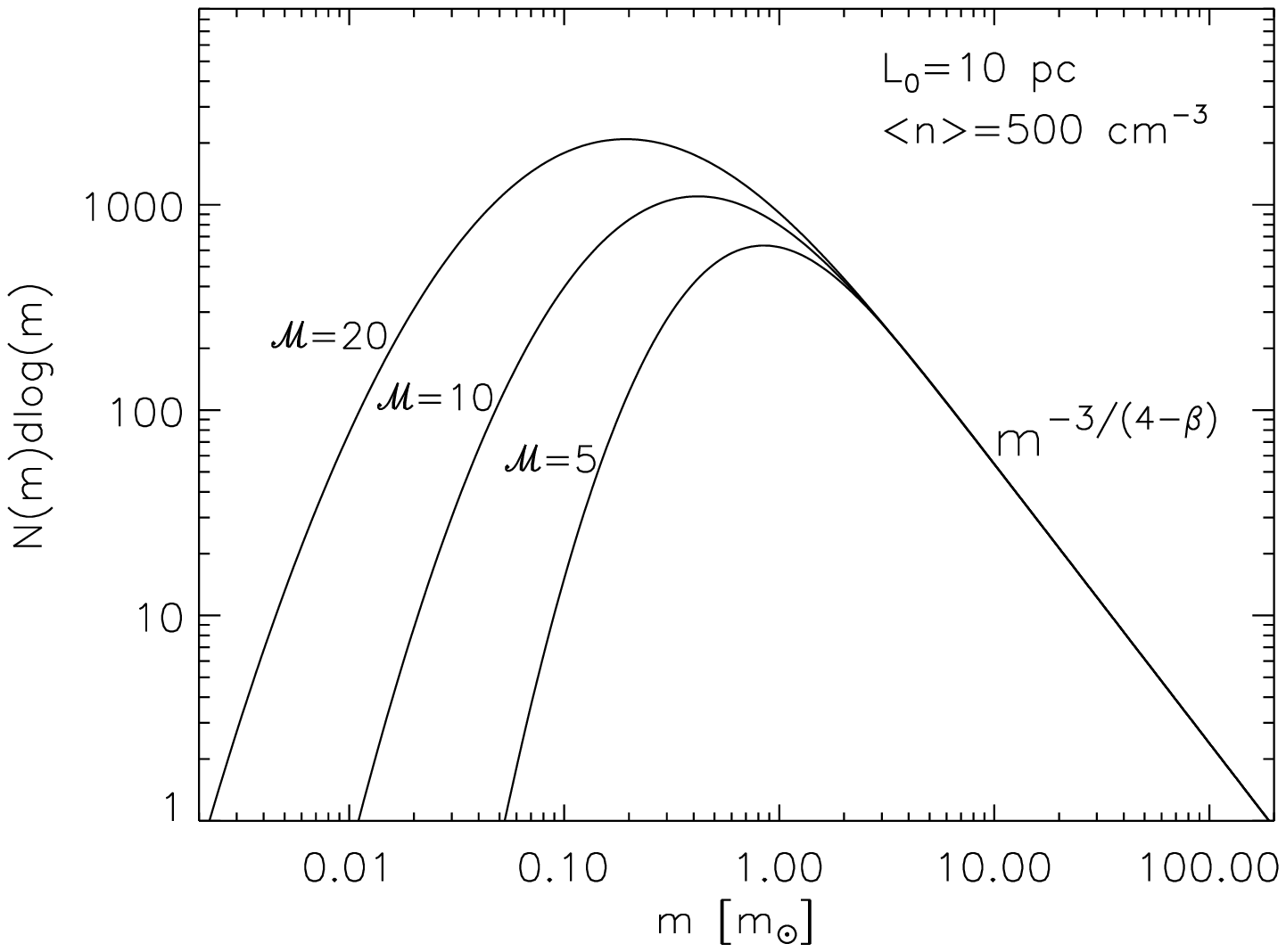}{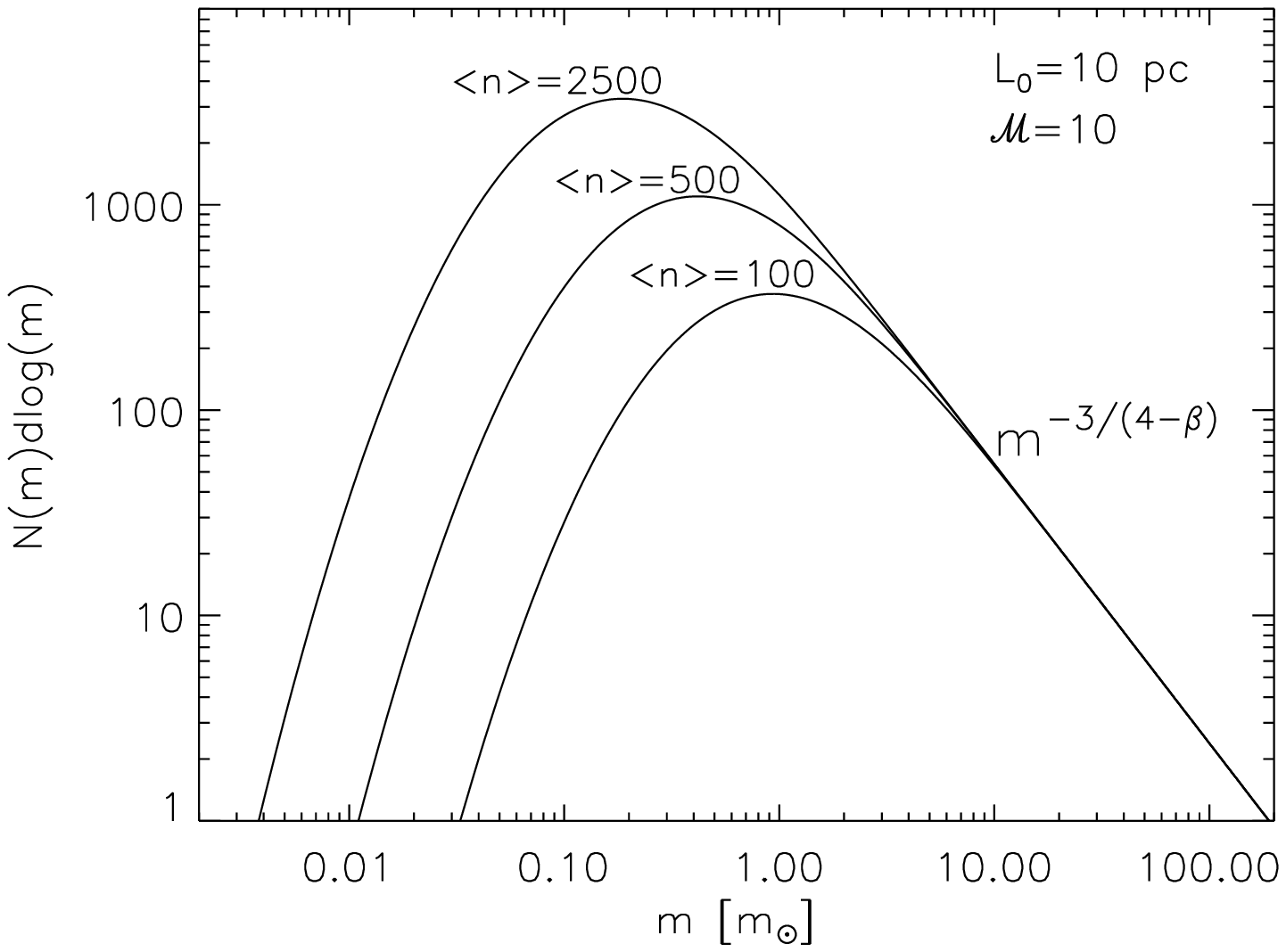}{.42}{Mass distribution of gravitationally
unstable cores from  equation (\ref{imfpdf}). Top panel: Mass distribution
as a function of the largest turbulent scale $L_0$, assuming
Larson type relations (for rescaling $\langle n\rangle$ and
${\cal  M}$ with $L_0$),  $T_0=10$~K and $\beta=1.8$.  Middle panel:
Mass distribution as a function of the rms Mach
number of the flow, assuming $\langle n\rangle=500$~cm$^{-3}$,
$T_0=10$~K and $\beta=1.8$. Bottom panel: Mass distribution as a function
of $\langle n\rangle$, assuming ${\cal M}=10$, $T_0=10$~K
and $\beta=1.8$.}

If $L_0$ is defined as the scale of a molecular cloud, with
average mass density $\rho_0$ and Alfv{\'e}nic Mach number
${\Mach}_{0}$, the mass of the largest cores formed by
turbulent fragmentation is estimated to be
\begin{equation}
m_{max}\approx \frac{\rho_0 L_0^3}{{\Mach}_{0}^2} .
\label{mmax}
\end{equation}
In MCs with mass $M_0\approx\rho_0 L_0^3\approx10^4$~M$_{\odot}$
and Mach number ${\Mach}_{0}\approx10$, $m_{max}\approx 100$~M$_{\odot}$.

While the majority of massive cores are larger than their Jeans' mass,
$m_J$, 
the probability that small cores are dense enough to collapse is
determined by the PDF of the density of the cores, which is
approximately Log-Normal. Even very small (sub-stellar) cores
have a finite chance to be dense enough to collapse.
If $p(m_J)\,d m_J$ is the Jeans' mass distribution obtained from the
PDF of gas density \cite{1997MNRAS.288..145P}, the fraction of cores
of  mass $m$ with gravitational energy in excess of their thermal
energy is given by the integral of $p(m_J)$ from 0 to
$m$. The mass distribution of collapsing cores is therefore
\begin{equation}
N(m)\, {\rm d}\log m\propto m^{-3/(4-\beta)}\left[\int_0^m{p(m_J){\rm d}m_J}\right]\,{\rm d}\log m ~.
\label{imfpdf}
\end{equation}
The mass distribution is plotted in \Fig{IMF}, for $\beta=1.8$.  In the
top panel the mass distribution is shown for three different values
of the  largest turbulent scale $L_0$, assuming Larson type relations
\cite{1981MNRAS.194..809L} to rescale the average gas density,
$\langle n\rangle$, and the
rms Mach number, ${\Mach}$, as a function of size,
$L_0$.  The mass distribution is a power law, determined by
the power spectrum of turbulence, for masses larger than approximately
1 m$_{\odot}$.  At smaller masses the mass distribution flattens,
reaches a  maximum at a fraction of a solar mass, and then decreases
with decreasing stellar mass. The mass distribution peaks at approximately
$0.4$~m$_{\odot}$ for the values ${\Mach}=10$,
$\langle n\rangle=500$~cm$^{-3}$, $T_0=10$~K and $\beta=1.8$,
typical of nearby molecular clouds. Collapsing sub-stellar masses are
found, thanks to the intermittent density distribution in the
turbulent flow. This provides a natural explanation for the origin
of brown dwarfs.

Note that the power law shape of the IMF for
mass values larger than about 1 m$_{\odot}$ is not affected by
the average physical properties of the system. On the
other hand the abundance of brown dwarfs is very sensitive to the
average gas density and the rms Mach number of the flow.
The  middle and  bottom panels of \Fig{IMF}  show the
dependence of the mass distribution on the rms Mach number of the flow
and on the average gas density respectively.
One can see in the middle panel that for an average gas density of
$\langle n\rangle=500$~cm$^{-3}$ and an rms Mach number ${\Mach}=5$,
typical of a molecular cloud complex such as Taurus, brown dwarfs
are very rare, while for the same average gas density and an rms
Mach number ${\Mach}=10$, typical of a molecular cloud complex such
as Orion (the density may be even larger), brown dwarfs are very
abundant (even more abundant if the IMF were plotted in units of
linear mass interval). This prediction is in fact unambiguously
confirmed by the observations \cite{2000ApJ...544.1044L}.

The thermal Jeans' mass is a more strict condition for
collapse than the magnetic critical mass. The magnetic critical
mass depends on the core morphology in relation to the magnetic
field geometry and strength. The latter correlates with the gas
density with a very large scatter \cite{1999ApJ...526..279P}.
It is possible therefore that magnetic pressure support against
the gravitational collapse limits the efficiency of star formation,
while its effect on the shape of the mass distribution is of secondary
importance.

Observations show that the stellar IMF is a power law above 1--2 m$_{\odot}$,
with exponent around the Salpeter value $x=1.35$, roughly independent
of environment \cite{Elmegreen_rev98,Elmegreen_rev2000},
gradually flattens at smaller masses, and peaks at approximately
0.2--0.6 m$_{\odot}$ \cite{%
1997AJ....113.1733H,%
1998A&A...336..490B,%
1999ApJ...525..440L,%
1999ApJ...525..466L,%
2000ApJ...540.1016L,%
2000ApJ...544.1044L}.
The shape of the IMF below 1--2 m$_{\odot}$, and
particularly the relative abundance of brown dwarfs, may depend on the
physical environment \cite{2000ApJ...544.1044L}. These observational
results are all consistent with our theoretical IMF.

It has been argued that only a small fraction of the mass of each
collapsing core may end up into the final star, due to mass loss
in protostellar winds, with a major effect on the stellar IMF.
However, stellar winds could be important for the origin of the stellar
IMF only if the ratio of initial core mass to final stellar mass were
comparable to the total mass range for stars ($\sim 10^4$, from
$\sim 100$~M$_{\odot}$ to $\sim 0.01$~M$_{\odot}$), as pointed out by
Elmegreen \cite{1999ApJ...522..915E}. This is highly unlikely,
because i) the correct slope and mass range of the IMF is already achieved
by turbulent fragmentation alone and ii) observational results indicate that
the mass distribution of prestellar cores is indistinguishable from the
stellar IMF \cite{%
1998A&A...336..150M,%
2001A&A...372L..41M,%
1998ApJ...508L..91T,%
1999sf99.proc..153O},
as predicted in earlier work on turbulent fragmentation and the origin of
the stellar IMF \cite{1995MNRAS.277..377P}.

\subsection{Mass distribution of prestellar cores in numerical simulations}

\FIG{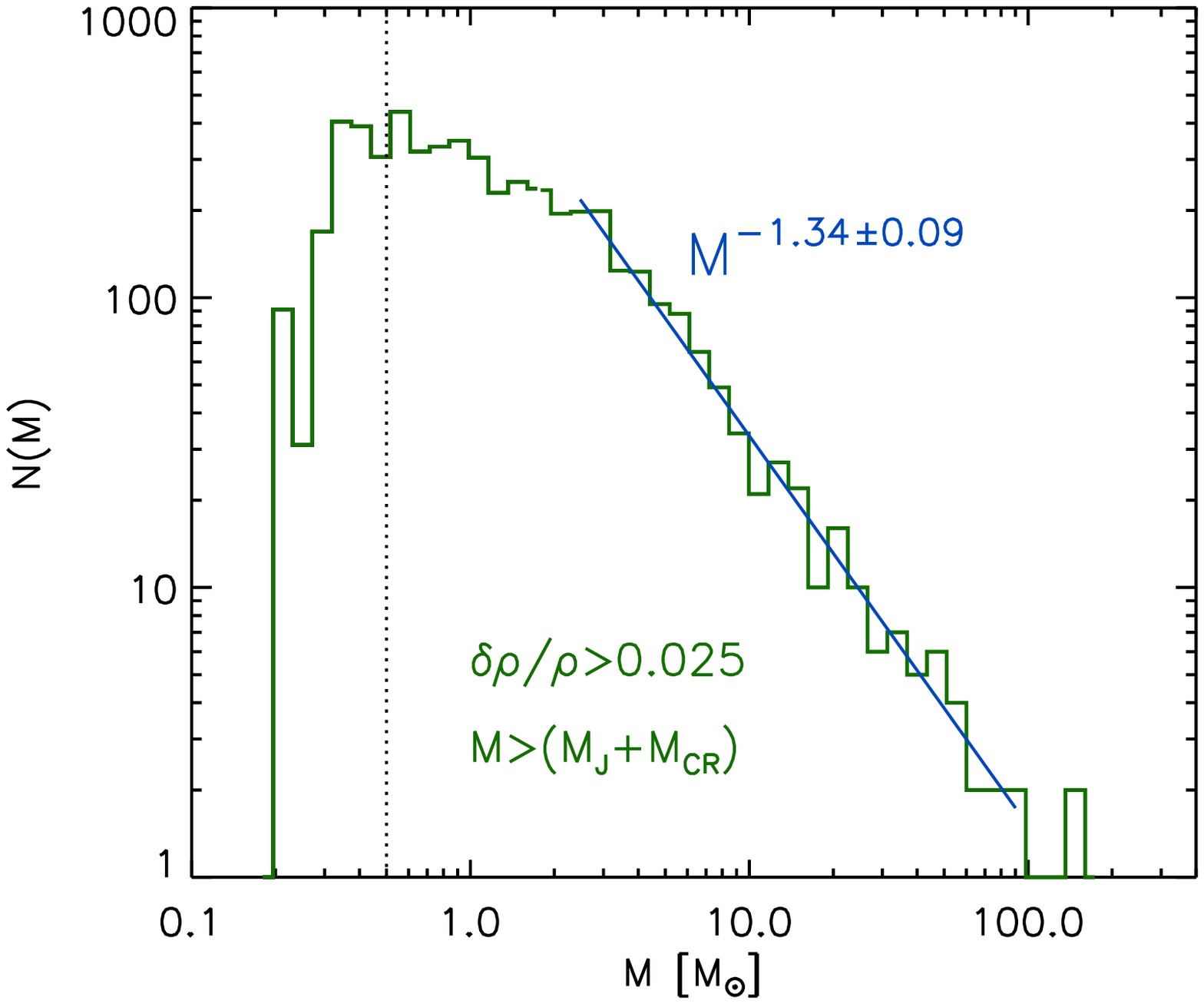}{.7}{Mass distribution of collapsing cores in the
range 0.2--100~M$_{\odot}$, derived from a $128^3$ numerical simulation.}

The mass distribution of prestellar cores may be measured directly
in numerical simulations of supersonic turbulence. With a mesh of
250$^3$ computational cells, and assuming a size of the simulated
region of a few pc, it is not possible to follow numerically the
gravitational collapse of individual protostellar cores. However,
dense cores at the verge of collapse can be selected in numerical
simulations by an appropriate clumpfind algorithm. Such an
algorithm should scan all density levels and recognize when
a large core is fragmented into smaller and denser ones, in which case
the large core should not be counted. Cores should also be excluded if their
gravitational energy is not large enough to overcome thermal and
magnetic support against the collapse, since only collapsing cores
should be selected.

\FIG{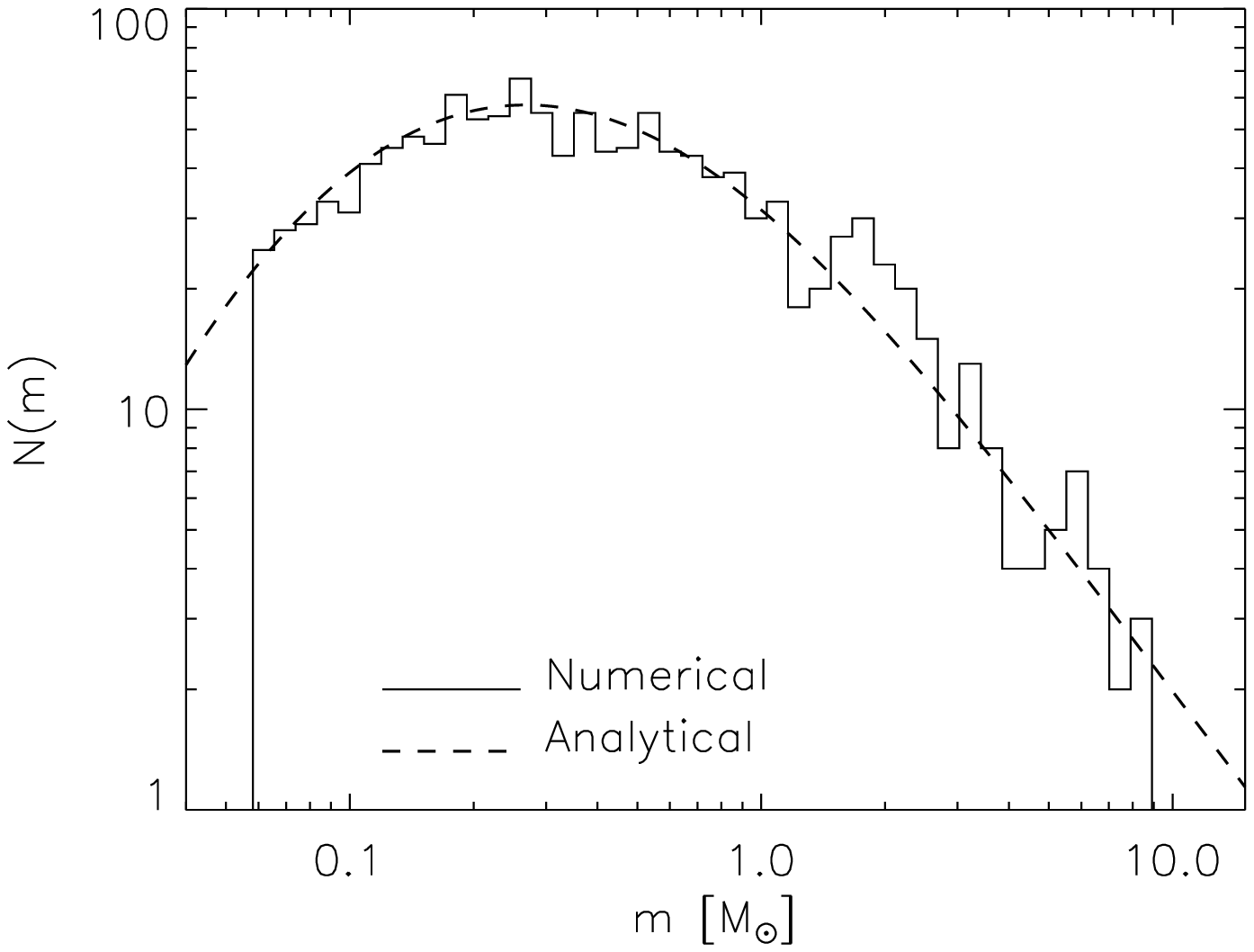}{.7}{Solid line: Mass distribution of collapsing cores,
derived from the density distribution of two snapshots of a 250$^3$
simulation with rms Mach number $\Mach\approx 10$.  The simulation is
scaled to physical units assuming $\langle n\rangle=500$~cm$^{-3}$,
$T_0=10$~K, and a size of 10~pc.  The fractional mass in collapsing
cores is 5\% of the total mass. Dashed line: Analytical mass
distribution, $N(m)$, computed for $\langle n\rangle=500$~cm$^{-3}$,
$T_0=10$~K, $\beta=1.8$.}

A mass distribution of collapsing cores, derived from the density
distribution in a numerical simulation is shown in \Fig{imf-num-128}.
The computational box with 128$^3$ cells has been scaled to two
scale ranges, suitable for
sampling cores in the intervals 0.2--2~M$_{\odot}$ and
2--100~M$_{\odot}$, respectively.  The mass distribution above
1~M$_{\odot}$ is a power law consistent with our analytical result and
with the observations. Below 1~M$_{\odot}$ the histogram flattens and
then turns around at approximately 0.3~M$_{\odot}$, also consistent
with the analytical theory and the observations. The cut-off at $\sim$
0.2~M$_{\odot}$ is due to the finite numerical resolution; the grid
size, rms Mach number, and mean density together impose a limitation on
the mass of collapsing cores.

Stretching the mass interval of sampled cores further into the brown
dwarf regime requires larger numerical resolution. \Figure{imf-num} shows
the mass distribution of collapsing cores derived from two snapshots of
a 250$^3$ simulation. The average gas density has been scaled to
500~cm$^{-3}$ and the size of the computational box to 10~pc. These
values have been chosen to be able to select cores in a range of masses
from a sub-stellar mass to approximately 10~M$_{\odot}$.  With this
particular values of average gas density and size of the computational
box, the smallest mass that can be achieved numerically is
0.057~M$_{\odot}$. Brown dwarfs masses ($<0.08$~M$_{\odot}$) are
therefore included. With an even larger numerical mesh, or assuming a
larger average density (and a smaller size), even smaller masses would
be selected. Turbulent fragmentation thus provides a natural
explanation for the origin of brown dwarfs. This was found from the
analytical model of the IMF presented above and is here confirmed from
the numerical mass distribution.

\FIG{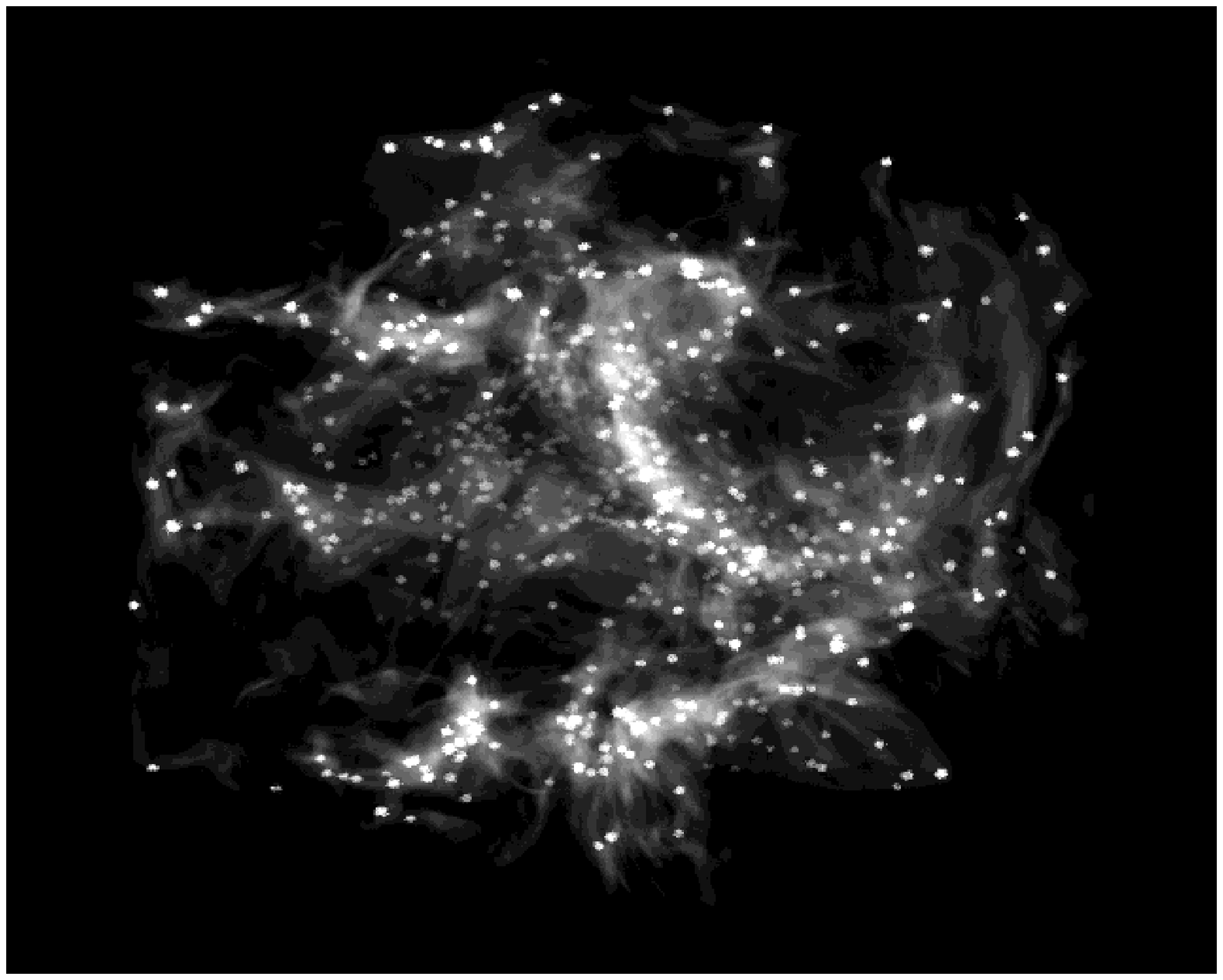}{.80}{Voxel projection of the density field of a snapshot
of a 250$^3$ numerical simulation of supersonic and super-Alfv\'{e}nic
turbulence. Collapsing cores are highlighted as bright spheres, with
brightness and size varying as a function of the core mass. The
brightness also depends on the column density of gas between each core
and the observer, in order to mimic the effect of dust extinction.
Slightly extended patches of bright emission are ``unresolved''
stellar clusters. The fraction of mass in collapsing cores is about
5\% in this simulation.}

Observed star-forming clouds appear very filamentary, as the projected
density field of supersonic turbulent flows. We have performed
accurate comparisons of statistical properties of turbulent flows
with observational data, by computing synthetic spectral maps of
molecular transitions \cite{%
1998ApJ...504..300P,%
1999ApJ...525..318P,%
2001ApJ...553..227P}.
The synthetic spectral maps are obtained by computing the non-LTE
radiative transfer problem using the density and velocity fields
of the MHD simulations. We have shown that fundamental
statistical properties of supersonic turbulence are unambiguously
found in the observational data of star-forming clouds
\cite{2001ApJ...553..227P}.

In star-forming clouds, prestellar cores and young stars tend to
concentrate in the densest filaments and cores. Since filaments
and cores of the same nature are found in the numerical simulations,
it is interesting to visualize the position of the collapsing cores
selected numerically, relative to the gas density distribution.
In \Fig{cloud} a voxel projection of the density field is shown,
where all the numerically selected cores have been highlighted as
bright spheres. The size and brightness is a function of the core
mass. The brightness is also a function of the optical depth of the
gas between the observer and the stars, to mimic the effect of
dust extinction. \Figure{cloud} shows beautiful filamentary structure
in both the gas and the stellar distribution, very reminiscent of
observed star-forming regions. It is quite amazing that a numerical
simulation of randomly driven supersonic and super-Alfv\'{e}nic
turbulence with periodic boundary conditions is able to produce
at the same time i) density structures morphologically and statistically
consistent with the observations; ii) prestellar cores correlated
with the gas distribution in a way qualitatively similar to the
observations and with a value of magnetic field strength typically
observed; iii) a mass distribution of the same prestellar cores
that agrees with the observed stellar (and prestellar cores) mass
distribution over the whole range of stellar masses, from brown
dwarfs to massive stars; and iv) a star formation efficiency
consistent with that in observed molecular clouds of only a few
percent per large scale dynamical time.

\section{Conclusions}
\label{concl.sec}

The main conclusion from the preceding sections is that the
statistics of star formation is primarily controlled by
supersonic turbulence, rather than by gravity.  Star formation
takes place in cold molecular clouds, which are part of a
turbulent cascade in the interstellar medium.  The ultimate
energy input to the cascade comes from supernovae, with a
possibly significant contribution from local variations of the
galactic rotation curve (density waves). The clouds owe their
existence to random convergence of the interstellar medium
velocity field, which creates local density enhancements over a
range of scales.  The internal, supersonic and super-Alfv{\'e}nic
velocity field in molecular clouds is responsible for their
fragmentation, thus preventing global collapse but
triggering local collapse at the many local density maxima whose
mass exceeds the local Jeans' mass. Such prestellar cores are
formed as sheet corrugations and filamentary density
enhancements, and are taken over by self-gravity only after they
have been shaped by the turbulence.

The velocity field of the cascade is dominated by power in
solenoidal (shearing) motions, even though it is supersonic
and super-Alfv{\'e}nic.  Its spectrum of kinetic energy is
less steep than its velocity and magnetic field power spectra,
which explains how conditions can be super-Alfv{\'e}nic on
small (molecular cloud) scales, even though there is rough
equipartition between magnetic and kinetic energy density
on large (disk thickness) scales.

A Salpeter like IMF is the result of the near-self-similar,
power law nature of turbulence in molecular clouds, in
combination with density jump amplitudes determined by
MHD-shock jump conditions.

Star formation (at least in our galaxy) bites its own tail;
it is driven by supernovae and at the same time the birth
of massive stars gives
rise to new supernovae that re-enforce the driving.  External
sources of turbulence, such as kinetic energy from galaxy
collisions and merging may be the primary driving agent
in star-burst galaxies.

Different physical conditions (primarily higher temperatures
and lower metal abundances) in the Early Universe would
lead to higher mass at the low-mass cut-off, and a much
weaker magnetic field would lead to a steeper IMF slope.

\section*{Acknowledgments}

The work of {\AA}N was supported by a grant from the Danish
Natural Science Research Council, and in part by the Danish National
Research Foundation, through its establishment of the Theoretical
Astrophysics Center. The work of PP was in part performed while
PP held a National Research Council Associateship Award at the
Jet Propulsion Laboratory, California Institute of Technology.


\newcommand{\aap}{Astron.\ Astrophys.}
\newcommand{\apj}{Astrophys.\ J.}
\newcommand{\apjl}{Astophys.\ J. Lett.}
\newcommand{\mnras}{MNRAS}
\newcommand{\prl}{Phys.\ Rev.\ Lett.}

\newcommand{\jtitle}[1]{}

}

\end{document}